\documentclass[aip,amsmath,amssymb,reprint,nofootinbib,
]{revtex4-2}

\usepackage{graphicx}% Include figure files
\usepackage{dcolumn}% Align table columns on decimal point
\usepackage{bm}% bold math
\usepackage{comment} 
\usepackage{ulem}
\usepackage[utf8]{inputenc}
\usepackage[T1]{fontenc}
\usepackage{mathptmx}
\usepackage{etoolbox}
\usepackage{color} 
\usepackage{hyperref}
\makeatletter
\def\@email#1#2{%
 \endgroup
 \patchcmd{\titleblock@produce}
  {\frontmatter@RRAPformat}
  {\frontmatter@RRAPformat{\produce@RRAP{*#1\href{mailto:#2}{#2}}}\frontmatter@RRAPformat}
  {}{}
}%
\makeatother
\begin{document}

\preprint{AIP/123-QED}

\title{Early-stage impact dynamics in %\textcolor{red}{(moderately)} 
dense suspensions of millimeter-sized particles}
% Force line breaks with \\
\author{Hirokazu Maruoka}
\affiliation{Center for Gravitational Physics and Quantum Information, Yukawa Institute for Theoretical Physics, Kyoto University, Kyoto 606-8502, Japan
}
 \affiliation{Okinawa Institute of Science and Technology,
Onna, Okinawa 904-0497, Japan
 }\thanks{Present address}%Lines break automatically or can be forced with \\
  \email{hirokazu.maruoka@oist.jp. hmaruoka1987@gmail.com}
\author{Hisao Hayakawa}%
\affiliation{Center for Gravitational Physics and Quantum Information, Yukawa Institute for Theoretical Physics, Kyoto University, Kyoto 606-8502, Japan
}
\email{hisao@yukawa.kyoto-u.ac.jp}%

\date{\today}% It is always \today, today,
\footnotetext{The following article has been submitted to Physics of Fluids. After it is published, it will be found at https://pubs.aip.org/aip/pof.}

\begin{abstract}

We investigate the early-stage impact dynamics in dense suspensions composed of millimeter-sized particles. 
While traditional models based on Stokes flow are typically applicable to suspensions of micrometer-sized particles, their validity for larger particles remains uncertain. 
Through controlled impact experiments, we examine the maximum drag force $F_{\mathrm{max}}$ acting on a projectile as a function of impact speed $u_0$. 
We have successfully conducted experiments using these
suspensions and confirmed the relation $F_\mathrm{max}\sim u_0^{3/2}$ for relatively large $u_0$ as observed in the previous studies of suspensions of micrometer-sized particles.
Our results demonstrate that the floating model—equivalent to the viscous model—successfully captures the early-stage dynamics even in suspensions with larger particles. 
This finding suggests that viscous-dominated behavior persists under certain conditions, extending the applicability of the floating model or the viscous model to new regimes. 
Our work provides experimental validation for theoretical predictions and contributes to a deeper understanding of impact-induced responses in dense suspensions.

\end{abstract}

\maketitle

\section{\label{sec:level1} Introduction }

Suspensions of colloidal particles in liquid can be commonly observed in nature~\cite{Russel89,Hunter01,Mewis11}. 
Knowing the flow properties of a suspension is important for both natural science and industry. 
%Although the viscosity of a Newtonian fluid is independent of the shear rate, there are many domestic substances (liquids containing microstructures such as suspensions and polymers) where the viscosity depends on the shear rate (non-Newtonian fluids). 
Some suspensions, as typical non-Newtonian fluids, exhibit discontinuous shear thickening (DST) in which the viscosity abruptly changes discontinuously from a small value to a large value at a critical shear rate. 
DST has attracted considerable attention among physicists in recent years ~\cite{Barnes89,Lootens05,Mewis11,Seto13,Cwalina14,Brown14,Guy15,Ness22} as a typical nonequilibrium discontinuous phase transition between a liquid-like phase and a solid-like phase. 
The DST is also expected to be important for industrial applications such as traction controls.
There are various spectacular aspects in addition to DST in the rheology of dense suspensions.
See, e.g., Refs.~\cite{Brady88,HayakawaIchiki,Boyer,Guazzelli,Suzuki19} for both experimental and theoretical aspects.

An intriguing phenomenon within dense suspensions is impact-induced hardening (IIH), enabling locomotion to stop liquids while causing sinking for individuals standing or walking~\cite{Brown14}. 
This is related to the impact dynamics of a projectile on a plate~\cite{Stronge,Louge02,Kuninaka04,Falcon, Chastel, Maruoka}, or a solid particle~\cite{Stronge,Kuwabara87,Brilliantov96,Kuninaka09,Saitoh10,Muller13}, or granular beds~\cite{Uehara03,Walsh03,Massimo04,Lohse04, Katsuragi07, Katsuragi16,Krizou20}, but the impact dynamics in suspensions is more drastic because IIH is related to a dynamical phase transition from a liquid to a solid.
This behavior finds practical utility in applications such as protective vests~\cite{Lee03}. 
Extensive experimental~\cite{Waitukaitis12,Waitukaitis13,Han16,Roche13,Maharjan18,Egawa19,Brassard21} and theoretical~\cite{Pradipto21a,Pradipto21b,Pradipto23} studies have explored IIH including oscillation and stick-slip processes~\cite{Kann11} and relaxation process after the impact~\cite{Maharjan17,Cho22,Barik22}.
They found that IIH is caused by an elastic contact force between suspended particles, mostly observed only in suspensions of frictional particles confined in a shallow container~\cite{Egawa19,Pradipto21b}.
Some papers confirmed the existence of the dynamical jammed region in front of the projectile~\cite{Waitukaitis12,Waitukaitis13,Han16,Pradipto23}. 
They also confirmed that there is a viscous regime in an early stage of the impact process before the reaching the maximum drag force acting on the projectile $F_\mathrm{max}$ at time $t_\mathrm{max}$~\cite{Brassard21,Pradipto21b,Pradipto23}, and subsequently, an elastic force plays an important role~\cite{Egawa19,Pradipto21a,Pradipto21b,Pradipto23}.
Although IIH is similar to DST, IIH can be distinguished from DST~\cite{Brown14,Otsuki11,Seto13,Ness22} by its localized nature versus the global manifestation of DST, transient behavior versus steadiness, and distinct shear stress responses~\cite{Pradipto21a}.

A primary focus in impact dynamics research involves elucidating the impact experiments of projectiles\cite{Louge02, Falcon,Chastel,Maruoka}, where the time evolution of projectiles can be observed to determine the force based on contact mechanics\cite{Johnson}. 
%Early-stage impact dynamics in dense suspensions is governed initially by a viscous force~\cite{Brassard21,Pradipto21b,Pradipto23} 
Notably, empirical observations suggest a relationship between the maximum drag force $F_\mathrm{max}$ and impact speed $u_0$, approximated by $F_\mathrm{max}\sim u_0^{3/2}$ predicted by the viscous model~\cite{Brassard21} and the floating model~\cite{Pradipto21b,Pradipto23}, which are equivalent models.
This result is similar to the relation $F_\mathrm{max}\sim u_0^{4/3}$ observed for impact processes in dry granular materials~\cite{Krizou20,Mandal24}.

The previous numerical studies~\cite{Pradipto21a,Pradipto21b,Pradipto23} rely on a lattice-Boltzmann method (LBM) and a discrete element method.
Although their model analysis agrees with the experimental results~\cite{Brassard21}, several questions remain about the applicability of their analysis.
Their approach is believed to be valid only if the suspended particles immersed in a viscous fluid are sufficiently small that the Reynolds number is quite low and the Stokes approximation is valid.
Due to the limitations of their computer resources, their simulation contains fewer than 3000 suspension particles in fluids, and the radius of the projectile is only a few times larger than that of the suspension particles.
If suspended particles are a typical suspension, such as cornstarch or potato starch, the projectile must be quite small.
Increasing the size of the projectile while maintaining a constant size ratio between the suspension particles and the projectile may invalidate the Stokes approximation at the millimeter scale. 
This observation highlights the importance of the size of the dispersed particles in understanding impact processes at this scale.

Building upon prior investigation, this study experimentally demonstrates how the size of dispersed particles influences the early-stage dynamics of the impact processes in dense suspensions. 
Our exploration delves deeper into the mechanical response of suspension materials to sudden mechanical stimuli, uncovering the fundamental physical mechanisms. %driving the observed hardening phenomenon of the early stage. 
We critically reassess the applicability of theoretical analyses utilizing the floating model and LBM within our experimental framework, mainly focusing on early-stage dynamics and the relationship between $F_\mathrm{max}$ and $u_0$.
 Our findings affirm that the floating model or the viscous model, typically valid under Stokes flow conditions, holds relevance even for dense suspensions containing millimeter-sized particles.
The dramatic IIH process in the impact dynamics caused by an elastic response of suspensions is not the target of this study, but will be the subject in the proceeding research. 
%Our experimental results suggest that the floating model valid only in the Stokes flow can be used even for dense suspensions of millimeter-sized particles.

%\textcolor{red}{Please write the construction of this paper.}

The rest of the paper is structured as follows. 
In the next section, we depict the experimental setup. 
We introduce the theoretical framework, the floating model in detail in Sec.~\ref{Float}. 
We explain the experimental results to clarify the applicability of the floating model in Sec.~\ref{results}.
In Sec.~\ref{sec:discussion} we discuss our results.
We conclude our results with some remarks in Sec. \ref{sec:conclusions}.
In Appendix~\ref{appendix_a}, we clarify the sidewall effect used in the main text from the comparison of the experimental result with a cube container.
In Appendix~\ref{appendix_b}, we examine whether the simplified buoyancy force can be used to describe experimental results.
In Appendix~\ref{appendix_c}, we present the exact solution of the floating model.
In Appendix~\ref{appendix_d}, we examine how the experimental results depend on the projectile diameter and the viscosity of the solvent.
In Appendix~\ref{appendix_e}, we present a phenomenology to derive the effective viscosity for dense suspensions.

\section{Experiment}\label{sec:exp}
%{\it Experiment}--.
%%%%%%%%%%%%%%%%%%%%%%%
%%%%%%%%%%%%%%%%%%%%%%%%%%%%%%%%%%%%%%%%%%%%%% Keep this line to distinguish the figure environment from the main text %%%%%%%%%% 
\begin{figure}
\includegraphics[width=9cm]{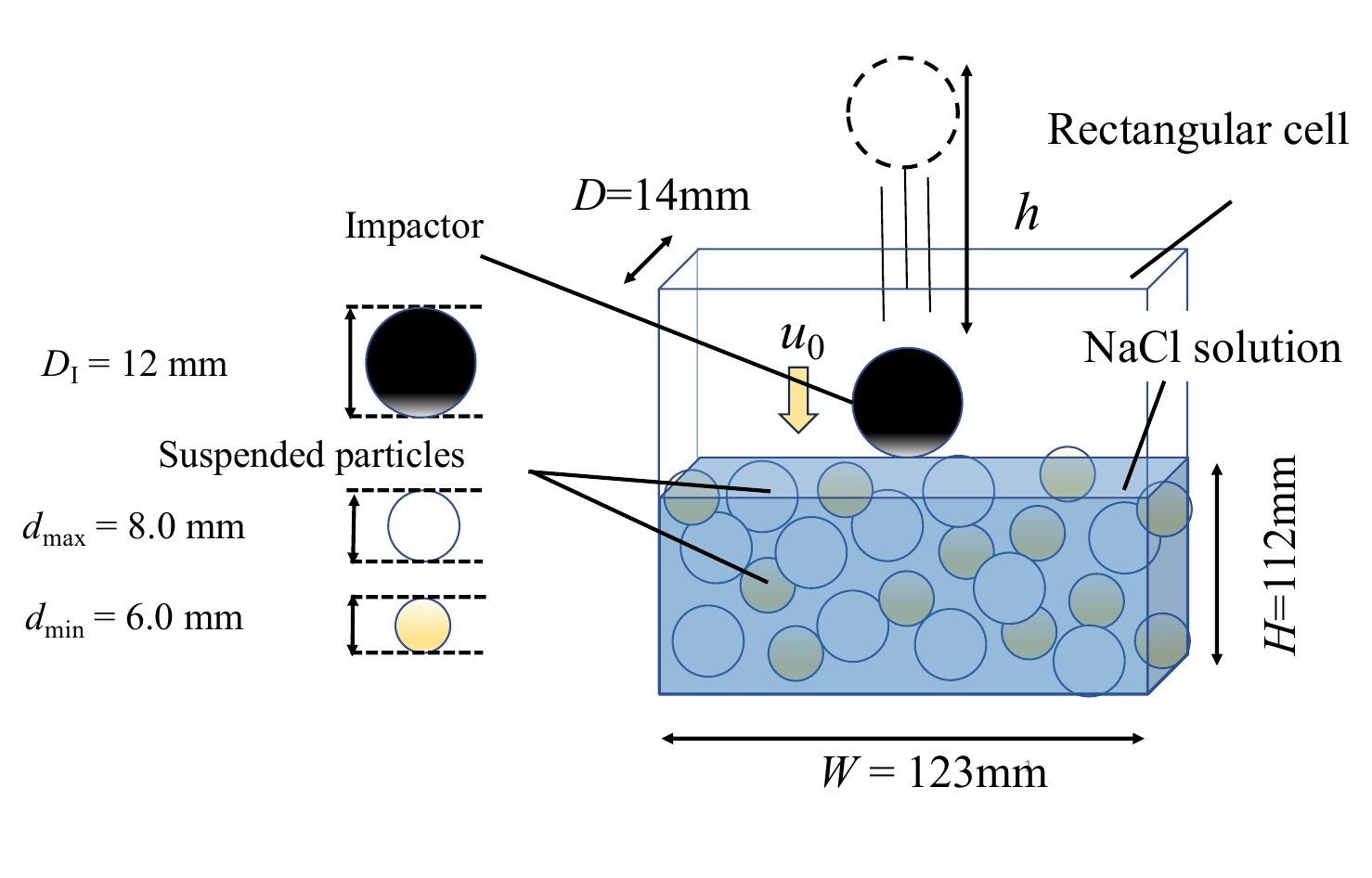}
\caption{\label{fig:F1} 
An illustration of our experimental setup. 
We prepare a suspension containing millimeter-sized bi-disperse particles (6.0, 8.0~{\rm mm}) in a solvent of NaCl solution confined in a quasi-two-dimensional container.
The density of the solvent is matched with that of the particles. 
By dropping a metallic spherical projectile from the heights $h$, the projectile collides with the suspension liquid with the impact speed $u_0$.
We record the impact processes in a high-speed camera.
%\textcolor{red}{The fonts of the figure must be updated.
%Do not rewrite the caption inappropriately! Your tense is wrong. Many other mistakes were inserted after my writing.}
}
\end{figure}
%%%%%%%%%%%%%%%%%%%%%%%%%%%%%%%%%%%%%%%%%%%%%%%%%%%%

Let us explain the setup of our experiment. 
We prepare a dense suspension by mixing millimeter-sized bi-disperse particles, consisting of two types of airsoft pellets: one with a maximum diameter of $d_{\rm max} = 8.0$ mm and a density of $1.03~{\rm g/cm^3}$ (MARUSHIN KOGYO CO.), and the other with a minimum diameter of $d_{\rm min} = 6.0$ mm and a density of $1.04~{\rm g/cm^3}$ (Tokyo Marui Co.), in a NaCl solution in the main text whose density is nearly matched with that of the dispersed particles ($\rho_{\rm f} \approx 1.05~{\rm g/cm^3}$).
%\textcolor{red}{The densities of particles are quite large (3-4$\%$ larger than the water). Are they correct?}
To observe the impact processes, we primarily use suspensions confined in a quasi-two-dimensional container ( Kenis, Ltd.) with the width $W = 123 {\rm mm}$, height $H = 112 {\rm mm}$, and depth $D = 14 {\rm mm}$.
See Fig.~\ref{fig:F1} as an illustration of the experimental setup.
As shown in Appendix~\ref{appendix_a}, we also examine a cube system with $W=D=94$mm, $H=90$mm to clarify the boundary effects in the quasi-two-dimensional container.
Then, we have confirmed that the results depend little on the side boundaries or box shapes.  

%\textcolor{red}{"Okuyuki" and "Haba" are the depth and width, respectively, in English. Thus, I have switched your wrong notations.}
We introduce the designed packing fraction defined as
$\varphi:= V_{\rm p} /V_{\rm s}$ to characterize the density of $N$ number of each particle,
where $V_{\rm p}: = \frac{4}{3} \pi N \left[ \left(d_{\rm max}/2 \right)^2 + \left(d_{\rm min}/2 \right)^2  \right]$ and  $V_{\rm s}:=W \times D \times H$. 
We prepare the suspension liquid in four different designed packing fractions $\varphi = 0.40, 0.48, 0.51,$ and $0.56$, which correspond to the number of particles $N=243,291,310,$ and $338$ respectively. 
Note that $\varphi=0.56$ is the maximum possible packing fraction as the suspension is no longer slurry \cite{Mitarai} if we study denser situations.
Note that particles are not uniformly distributed, and most of them tend to form a densely packed cluster due to imperfect density matching. 
We use the effective packing fraction $\Phi$ in the floating cluster near the surface of water for our data analysis in Sec. \ref{results}, where $\varphi=0.40, 0.48, 0.51$, and $0.56$ correspond to $\Phi=0.52, 0.53, 0.54$, and $0.56$, respectively. 
Although the dynamical jammed region produced by an impact becomes denser than the initial $\Phi$, the amount of compactification is not large~\cite{Pradipto23}.
Also, the experimental determination of the packing fraction in the dynamical jammed region is difficult.
Therefore, we use only $\Phi$ for the analysis of experimental results.

We use a metallic sphere as the projectile (diameter $D_{\rm I} = 12$ mm, $\rho_{\rm I} = 7800~{\rm kg \cdot m^{-3}}$ in the main text), which is suspended by an electromagnet (ESCO Co., Ltd., EA984CM-1) above the suspension before impact. 
After switching off the magnetic force, the projectile is dropped into the suspension with an initial impact speed $u_0$. 
The impact process is recorded using a high-speed camera (Phantom V641, Phantom T1340, Vision Research) at a frame rate of 10,000 fps. 
The impact velocity is controlled by varying the release height of the projectile ($h = 4 \sim 720~{\rm mm}$). 
We conduct repeated impact experiments (10 to 16 times per height) to minimize specific responses from peculiar configurations of the suspended particles.

We extract the trajectory of the projectile by an optical tracking technique with the aid of the Open CV library of Python. 
We set the marker on the top of the projectile to get its position in each flame. 
We adopt the second-order central difference scheme to obtain the velocity and the acceleration of the projectile.
Subtracting the buoyancy force acting on the projectile from the equation of motion of the projectile, we can evaluate the drag force acting on the projectile.

\section{Floating model}\label{Float}

\begin{figure*}
\includegraphics[width=18cm]{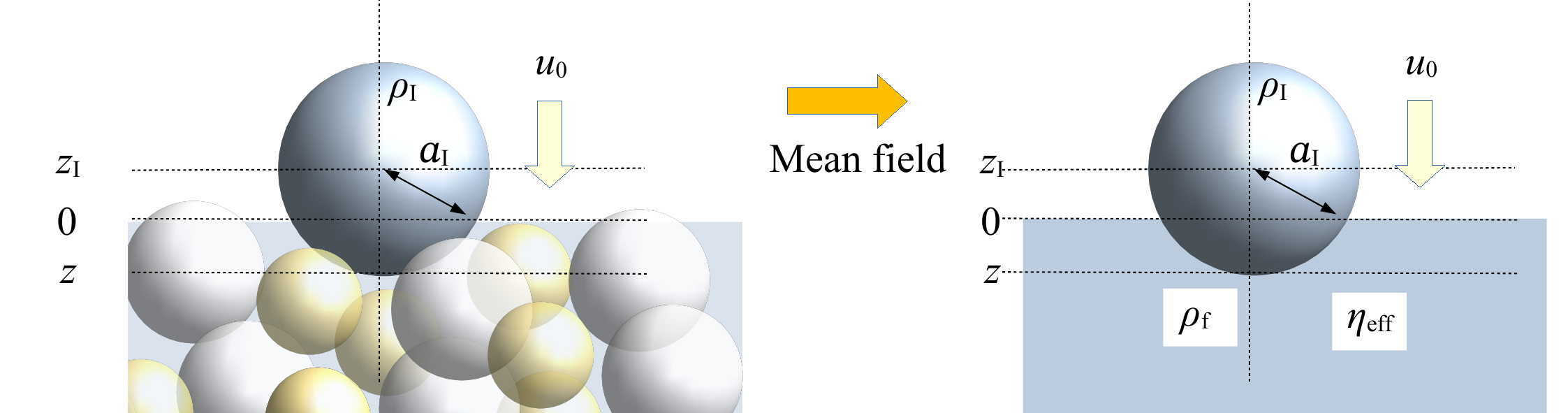}
\caption{\label{fig:F6} 
An illustration of a projectile in a suspension liquid, where the impact speed is $u_0$ when the projectile attaches to the surface of the liquid (Left).
We denote the position of the bottom head of the projectile $z(<0)$ where the center of mass is located at $z_{\rm I}$, the radius of the projectile $a_{\rm I}$, and the density of it is $\rho_{\rm I}$.  
The density of the solvent is $\rho_{\rm f}$. 
We introduce the effective viscosity $\eta_\mathrm{eff}$ acting on the projectile as a mean-field fluid of the suspension liquid in the floating model (Right).
%\textcolor{red}{Redraw the fonts in this figure.}
}
\end{figure*}

Brassard et al.~\cite{Brassard21} proposed the viscous model to describe the motion of the projectile in dense suspensions of micron-sized particles in the early stage of the impact dynamics.
Later, Pradipto and Hayakawa~\cite{Pradipto21b,Pradipto23} renamed it as the floating model, adding its derivation.\footnote{This name originates from floating suspended particles that do not have any elastic response even if a projectile hits suspended particles because they are floating.}
%Later, we have recognized the equivalency of the two models.
To get IIH, we need an elastic force~\cite{Roche13,Maharjan18,Egawa19,Pradipto21a,Pradipto21b,Pradipto23}, which appears only in a relatively late stage of the impact process, although the floating model involves only the viscous drag force acting on the projectile, except for the buoyancy force. 
In this paper, we adopt the floating model to describe the motion around $t_\mathrm{max}$ at which the drag force takes $F_\mathrm{max}$.
We depict an impact process in a suspension in the left figure of Fig.~\ref{fig:F6}, where the position of the bottom head of the projectile with radius $a_{\rm I}:=D_\mathrm{I}/2$ and its density $\rho_{\rm I}$ is denoted by $z$ and $z=0$ is fixed on the surface of the suspension.
Thus, we are only interested in the position $z$ satisfying $z\le 0$, i,e, under the influence of the suspension liquid.
%We denote the impact speed $u_0$ which corresponds to the speed of a projectile on the surface of the suspension liquid.
As a result of the drag force $F_{\rm D}$ and buoyancy force  $F_\mathrm{B}$ acting on the projectile, the projectile speed decreases with time and approaches zero.
We should note that the maximum force $F_\mathrm{max}$ acting on the projectile appears in the relatively early stage of impact processes, and the behavior of the projectile around $t_\mathrm{max}$ can be understood without the elastic force acting on the projectile, at least, for micro-meter sized suspensions and LBM simulations~\cite{Brassard21,Pradipto21b,Pradipto23}.

%%%%%%%%%%%%%%%%%%%%%%%%%%%%%%
%%%%%%%%%%%%%%%%%%%%%%%%%%%%%

Let us discuss the buoyancy force $F_\mathrm{B}$ acting on the projectile.
Although $F_\mathrm{B}$ should be proportional to the volume inside the liquid and the gravity directly acts on the projectile in the part outside the liquid, we simply adopt an approximate expression for $F_\mathrm{B}$ as~\cite{Pradipto21b,Pradipto23}
\begin{equation}
F_{\rm B} \approx  -m_{\rm I} g\left(1 - \frac{\rho_{\rm f}}{\rho_{\rm I}} \right) ,
\label{eq:e1}
\end{equation}
where $m_{\rm I}$ is the mass of the projectile, $g$ is the gravitational acceleration, $\rho_{\rm f}$ is the density of the solvent and $\rho_{\rm I}$ is the density of the projectile. The validity of this simplification is argued in Appendix~\ref{appendix_b}.
 
Next, we discuss the drag force $F_\mathrm{D}$ acting on the projectile.
In the floating model~\cite{Pradipto21b,Pradipto23} as well as the viscous model~\cite{Brassard21}, we assume that the drag force is proportional to the moving velocity.
This drag force is the result of a mean-field approximation of the suspension liquid as depicted in the right figure of Fig.~\ref{fig:F6}. 
Note that the derivation of the floating model can be found in Ref.~\cite{Pradipto21b}.  
If we assume such a model, the drag force acting on a partially filled projectile in the liquid is expressed as  
\begin{equation}
F_{\rm D}=  3 \pi \eta_{{\rm eff}} \frac{d z}{dt} z ,
\label{eq:e2}
\end{equation}
where $z$ is the vertical position of the head (deepest position) of the projectile, and $\eta_{\rm eff}$ is an effective viscosity of the mean-field fluid (see Fig.~\ref{fig:F6}). 
For simplicity, we assume that $\eta_\mathrm{eff}$ is independent of $u_0$ but depends on $\Phi$.
Although $\eta_\mathrm{eff}$ is the mean-field viscosity of the dynamical jammed region~\cite{Pradipto23} corresponding to the growth of the dynamical jammed region~\cite{Waitukaitis13}, it will be treated as a fitting parameter in the later analysis because of the difficulty of estimating the local viscosity. 
The expression of Eq.~\eqref{eq:e2} is reduced to the Stokes drag force for a filled projectile in the liquid for $z \le -2 a_{\rm I}$.
Thus, Eq.~\eqref{eq:e2} is only valid for $-2a_{\rm I}< z<0$.

The validity of the drag force in Eq.~\eqref{eq:e2} has already been confirmed for the impact dynamics in standard suspensions, such as cornstarch and potato-starch, and microscopic simulations~\cite{Brassard21,Pradipto21b,Pradipto23}.
It is, however, controversial whether the drag force $F_\mathrm{D}$ can be used for a suspension of millimeter-sized particles because fluid flows around a large particle to generate vortices.
Nevertheless, we may examine Eq.~\eqref{eq:e2} for a fluid flow around a projectile in a dense suspension of millimeter-sized particles because the interstitial distance between suspended particles is too small to generate vortices.
%In this paper, we adopt Eq.~\eqref{eq:e2} as the drag force acting on the projectile.
The validity of this model will be examined by comparing the model analysis and experimental results.

With the aid of Eqs.~\eqref{eq:e1} and \eqref{eq:e2}, the motion of the projectile in the early stage impact dynamics may be described by 
\begin{equation}
m_{\rm I}\frac{d^2 z}{dt^2}= -m_{\rm I} g \left(1 - \frac{\rho_{\rm f}}{\rho_{\rm I}} \right) + 3 \pi \eta_{{\rm eff}} \frac{d z}{dt}  z. 
\label{eq:e3}
\end{equation}
The dimensionless form of Eq.~\eqref{eq:e3} is expressed as
\begin{equation}
\frac{d^2 Z}{d \tau^2} = - \left(1-\xi \right) +  \eta \frac{d Z}{d \tau }Z ,
\label{eq:e4}
\end{equation}
where 
\begin{equation}
Z: = \frac{z}{a_{\rm I}},~\tau: =\sqrt{\frac{g}{a_{\rm I}}}t,~\xi:=\frac{\rho_{\rm f}}{\rho_{\rm I}},~\eta: = \frac{3 \pi \eta_{{\rm eff}} a_{\rm I}^{3/2}}{g^{1/2}m_{\rm I}},~U_0: = \frac{u_0}{\sqrt{a_{\rm I} g}}.
\label{eq:e5}
\end{equation}
Equation~\eqref{eq:e4}, which is equivalent to Eq.~\eqref{eq:e3}, is the equation of the floating model~\cite{Brassard21,Pradipto21b,Pradipto23}. 

As shown in Refs.~\cite{Pradipto21b,Pradipto23} and Appendix~\ref{appendix_c}, we can solve Eq.~\eqref{eq:e4} exactly. 
The exact solution of Eq.~\eqref{eq:e4} leads to the power-law relations of $F_\mathrm{max}$ %and $\tau_\mathrm{max}$ 
with the high impact speed $U_0$:~\cite{Pradipto21b}
\begin{equation}
\tilde{F}_\mathrm{max} \approx \frac{\sqrt{2 \eta}}{3} U_0^{\frac{3}{2}}, 
%\quad \tau_\mathrm{max} = \frac{1}{\sqrt{2}}U_0^{-1/2}
\qquad \left( \eta U_0 \gg 1 \right) 
\label{eq:e5a}
\end{equation}
where $\tilde{F}_\mathrm{max} := F_\mathrm{max}/m_{\rm I} g$.
%\textcolor{red}{Write the expression of $t_\mathrm{max}$ with the correct coefficient. 
%Do not hide disadvantageous results!}

\section{Results}\label{results}
%%%%%%%%%%%%%%%%%%%%%%%%%%%%
\begin{figure}
\includegraphics[width=9cm]{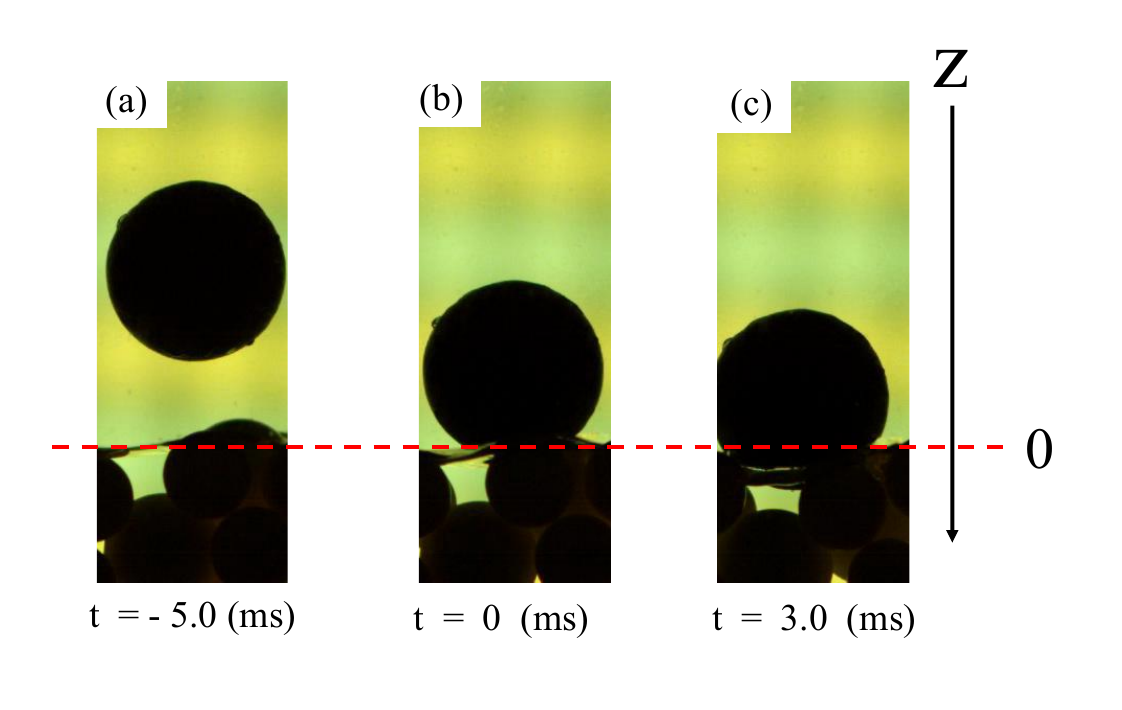}
\caption{\label{fig:F2} 
A set of experimental snapshots of an impact process for $\Phi=0.56$ and $u_0 = 1.6~{\rm m/s}$. 
Note that 3.0 ms after the first impact (the middle figure) the force acting on the projectile exhibits the maximum value $F_\mathrm{max}$ (Multimedia available online).
}
%After 2.2 ms from the first contact, the drag force is at its maximum $F_\mathrm{max}$. The movie is uploaded as Supplimental Materals.}
\end{figure}
%%%%%%%%%%%%%%%%%%%%%%%%%%%%%%%%%%%%%%%%%

%{\it Results}--.
Let us present the experimental results and examine whether the floating model introduced in the previous section can be used in the early stage of dynamics, focusing on the relationship between $F_\mathrm{max}$ and $u_0$.
Figure~\ref{fig:F2} is a typical set of experimental snapshots of an impact process for $\varphi=\Phi=0.56$ and $u_0 = 1.6~{\rm m/s}$ (Multimedia available online). 
The impact speed $u_0$ and the initiation time of the impact ($t=0$) are defined as the points of its maximum velocity and the corresponding time, respectively.
Then, the time $t$ begins with the initiation time.
Figure~\ref{fig:F2} (b) is a snapshot corresponding to the first hit of the projectile on the liquid surface at $t=0$.
Figure ~\ref{fig:F2} (c) is a snapshot when the drag force acting on the projectile reaches the maximum value $F_\mathrm{max}$ at $t_\mathrm{max}=3.0$ ms after the first impact.
%\textcolor{red}{

\begin{figure}
\includegraphics[width=9cm]{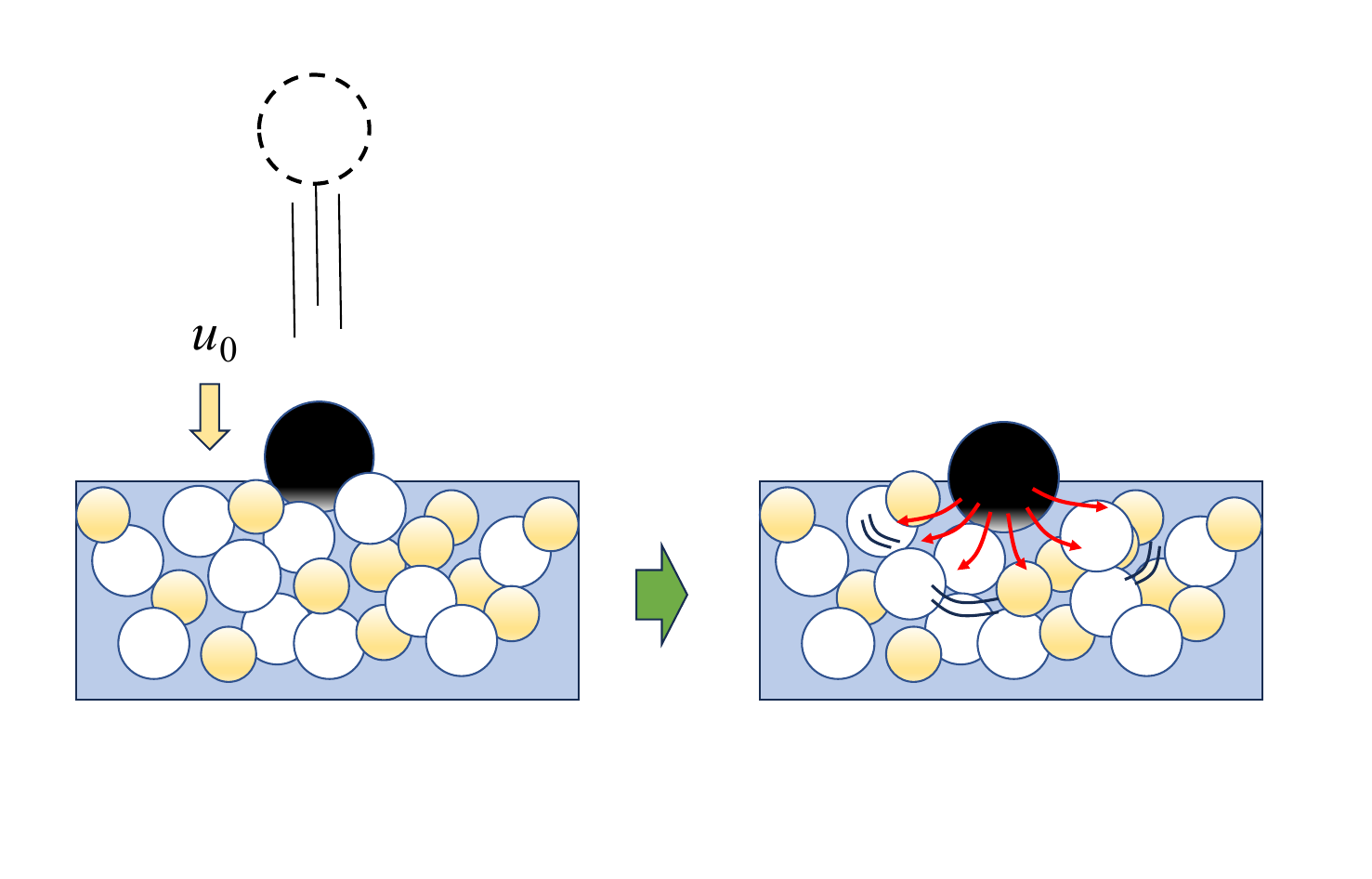}
\caption{The floating relaxation of dispersed particles after an impact, where a black ball represents the projectile. 
After the impact, the floating suspended particles are moved to relax into another stable position. 
%\textcolor{red}{図１個での説明は無理だと思います。}
\label{fig:FB1}
}
\end{figure}

Interestingly, most trials exhibit that the projectile hits one of the suspended particles at the first hit on the liquid surface.
This might cause some elastic response.
However, since the suspended particles are floating and particles are lubricated with the other particles, the hit of the projectile suppresses elastic response (Fig.~\ref{fig:FB1}).
If the suspended particles are percolated from the contacting particles with the projectile to the bottom plate, we can get a significant elastic response~\cite{Egawa19,Pradipto21b}.
However, this only appears in the late stage for $t>t_\mathrm{max}$.
Nevertheless, there are elastic effects caused by contact between the projectile and suspended particles that are not accounted for in the floating model.

%%%%%%%%%%%%%%%%%%%%%%%%%%%%%%%

%%%%%%%%%%%%%%%%%%%%%%%%%

%%%%%%%%%%%%%%%%%%%%%%%%%%%%%%%
\begin{figure}
\includegraphics[width=9cm]{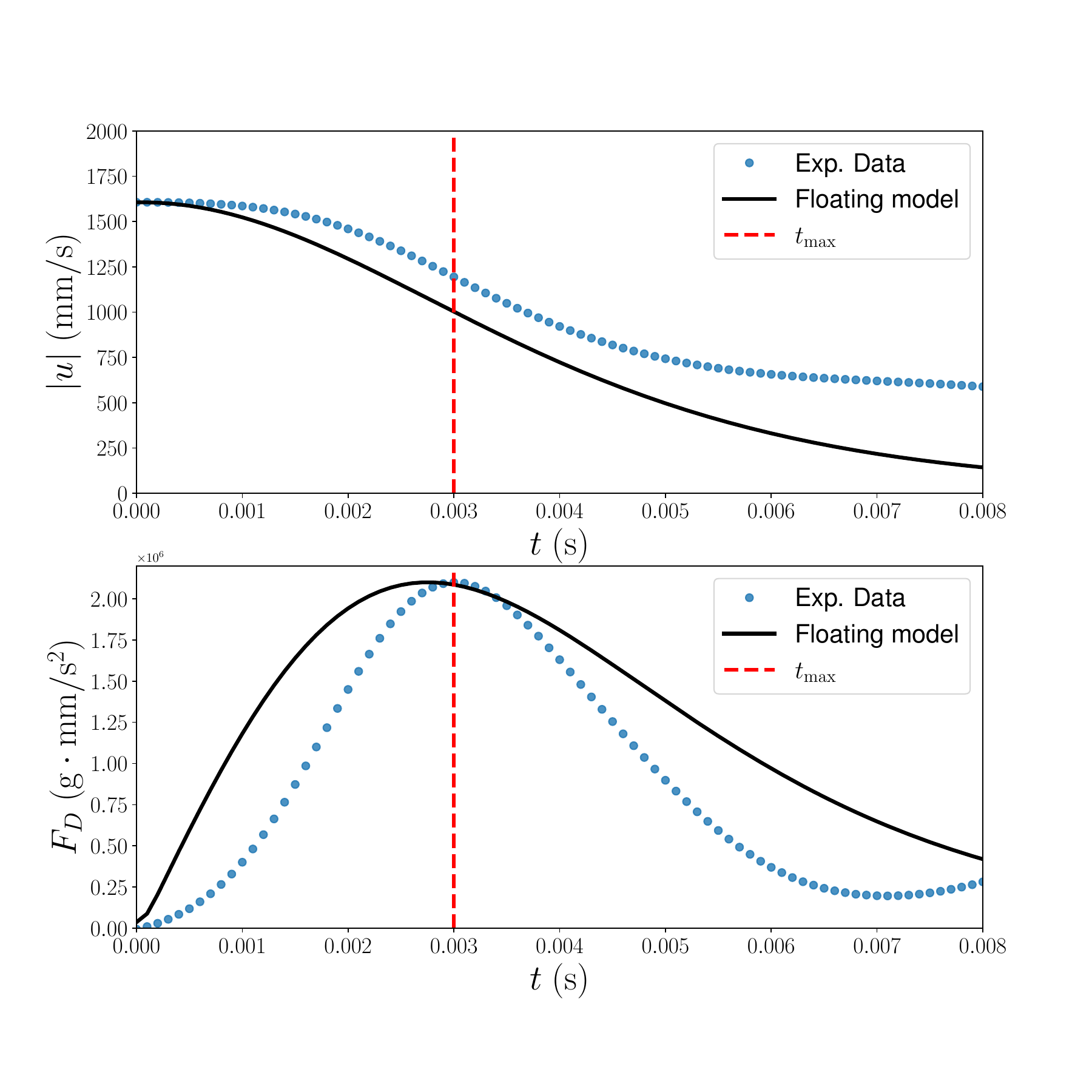}
\caption{\label{fig:F3} 
The time evolution of the velocity of the projectile (a) and the drag force acting on the projectile (b) for $\Phi=0.56$ and $u_0 = 1.6~{\rm m/s}$. 
The blue dots are experimental data. 
The solid lines are the solution of Eq.~\eqref{eq:e3} with a fitting parameter $\eta$. 
The dashed vertical red line indicates the time $t_\mathrm{max}$ to take $F_\mathrm{max}$. 
%\textcolor{red}{実験データですら$t_\mathrm{max}$でピークになっていない。(ピークは2点程左。)これはレフェリーも指摘していたが、改善がみられていない。}
}
\end{figure}
%%%%%%%%%%%%%%%%%%%%%%%%%%%%%%%%%%

The projectile stops in the middle of the suspension for the dense suspensions ($\Phi\ge 0.54$ corresponding to $\varphi\ge 0.51$), where the projectile cannot sneak into the suspension liquid for low-speed impact, and it can sneak into the liquid completely for high-speed impact. 
For lower packing fractions, such as $\Phi=0.52$ (corresponding to $\varphi=0.40$), the projectile penetrates the liquid completely and reaches the bottom of the container. 

%%%%%%%%%%%%%%%%%%%%%%
\begin{figure*}
\includegraphics[width=18cm]{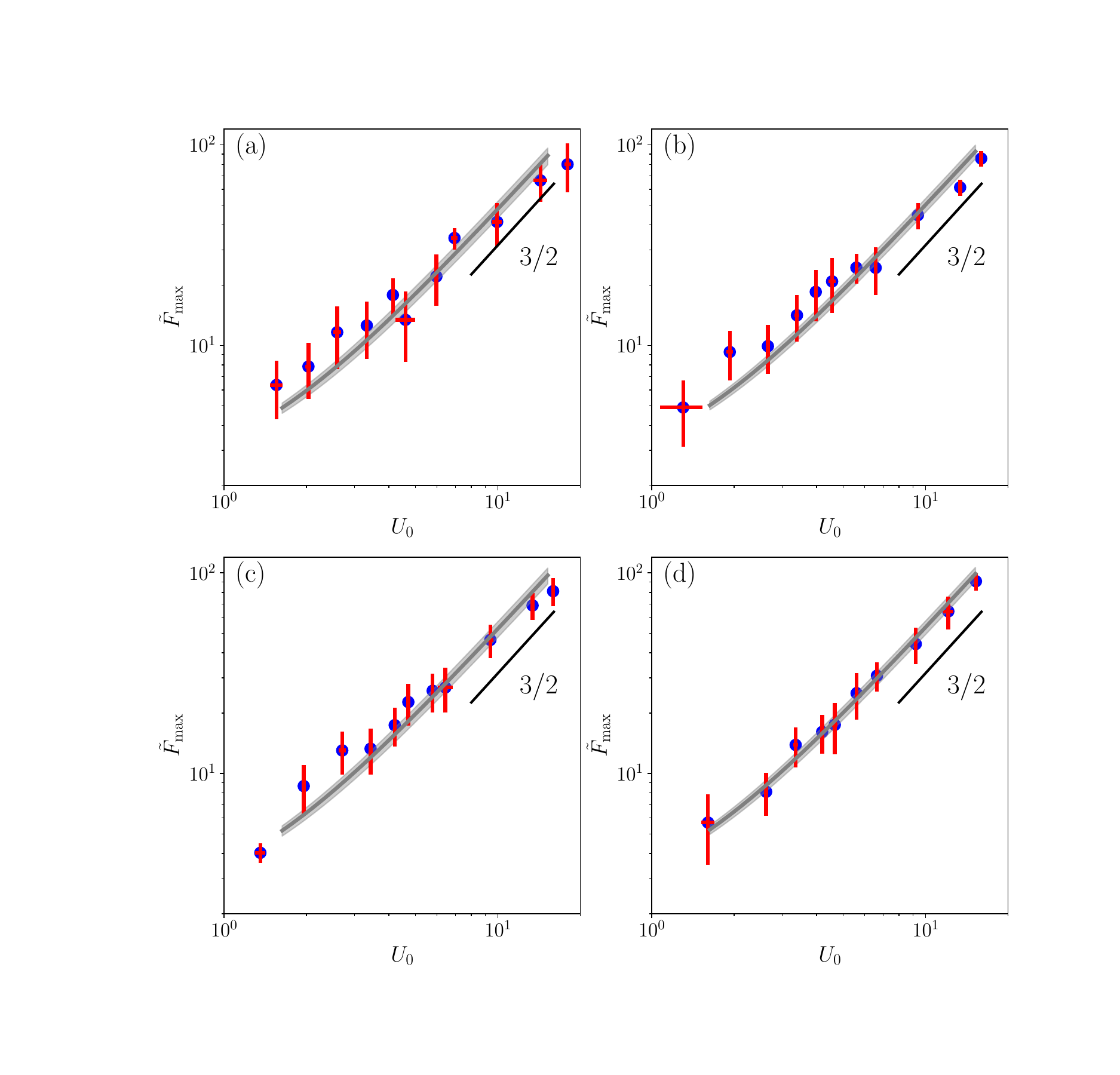}
\caption{The plots between the dimensionless maximum force $\tilde{F}_\mathrm{max}:=F_\mathrm{max}/m_{\rm I}g$ and impact-speed $U_0:=u_0/ \sqrt{a_{\rm I} g}$ for various packing fractions $\Phi=0.52$ (a), 0.53 (b), 0.54 (c), and 0.56 (d) corresponding to $\varphi=0.40, 0.48, 0.51$, and $0.56$, respectively. 
The blue solid circles indicate the experimental results, and the dark-gray solid lines are obtained by the numerical solutions of the floating model Eq.~\eqref{eq:e4}. The gray area around the solid line indicates the possible solutions within the errors of the estimated $\eta$.  
The black solid line segments in large $U_0$ are the guide lines $\tilde{F}_\mathrm{max}\propto U_0^{3/2}$ as expected from Eq.~\eqref{eq:e5a}.
%\textcolor{red}{LabelやLegendの字が変.pngを使っているせいだと思う。pngは使わないでepsかpdfにして下さい。(c)の図に変な白い四角がある。}
\label{fig:F4}}
\end{figure*}
%%%%%%%%%%%%%%%%%%%%%

%With the aid of optical tracking, we can trace the motion of the projectile.
As explained in Sec.~\ref{sec:exp} we can trace the velocity and acceleration of the projectile.
By using Eq.~\eqref{eq:e1} we can evaluate the drag force acting on the projectile.
Figures~\ref{fig:F3} depict the time evolutions of the velocity and the drag force acting on the projectile. 
%The drag force is estimated by subtracting the buoyancy force $F_\mathrm{B}$ from the acceleration of the projectile. 
As shown in Fig.~\ref{fig:F3} the projectile slows down after the impact by the drag force $F_D$ where $F_D$ takes the maximum value $F_\mathrm{max}$ at $t_\mathrm{max}$.
In these figures, the dots are obtained in the experiments, and the solid lines are the fitting curves using the solution of Eq.\eqref{eq:e3}. 
We choose $\eta:=3 \pi \eta_{{\rm eff}} a_{\rm I}^{3/2}/(g^{1/2}m_{\rm I})$ as a fitting parameter to recover experimental $\tilde{F}_\mathrm{max}$ in the floating model.
%The fitting is performed by obtaining $\eta$ to give the observed in the experiment.  
%\textcolor{blue}{
The behavior of the experimental data is qualitatively similar to that obtained from the floating model, although the very early stage dynamics of the projectile observed in the experiment seem to differ from the prediction of the floating model.
Therefore, we cannot expect quantitative agreement between theoretical $t_\mathrm{max}$ and experimental $t_\mathrm{max}$ as shown in Fig.~\ref{fig:F3}.
This disagreement between the theory and the experiment differs from the agreement between the LBM simulation and the theory~\cite{Pradipto21b, Pradipto23}.
We also stress that the experimental data exhibit different time evolution from that of the floating model for $t>t_\mathrm{max}$. 
%}
Here, we focus only on the early stage dynamics for $t\le t_\mathrm{max}$ in this paper; the late stage falls outside the scope of the present study.
%Thus, this result suggests that the floating model is valid to describe the motion of the projectile in the early stage,  for $t\le t_\mathrm{max}$, $u_0=3.1$ m/s, and $\Phi=0.56$.

Figure \ref{fig:F4} plots of $\tilde{F}_\mathrm{max}$ against $U_0$ for various packing fractions, $\Phi=0.52, 0.53, 0.54$, and $0.56$. 
The blue solid circles indicate experimental estimations of $\tilde{F}_\mathrm{max}$, and the dark-gray solid lines are the numerical solutions of Eq.~\eqref{eq:e4}, where the dimensionless effective viscosity $\eta$ is treated as a fitting parameter with its error bar.  
The gray areas around the solid line indicate the possible solution within the error of the estimated $\eta$.
%\textcolor{red}{You may need more explanation about the gray zone.}
It is readily observed that the theoretical curves closely overlap with the experimental data points, including their error bars. 
This agreement is particularly pronounced at higher packing fractions ($\Phi = 0.54$ and $0.56$), although the theoretical predictions successfully capture the overall trend of the experimental data even at lower packing fractions ($\Phi = 0.52$ and $0.53$).
We also confirm the asymptotic behavior predicted from the floating model as $\tilde{F}_\mathrm{max}\sim U_0^{3/2}$ for $\eta U_0\gg 1$ in Eq.~\eqref{eq:e5a} \cite{Pradipto21b,Pradipto23}. 
The black solid line segments in Fig.~\ref{fig:F4} indicate the power exponents of the asymptotic solution, which agree with the slope from the experimental data points. 
These results indicate that the experimental data across the entire range from $\Phi = 0.52$ to $\Phi = 0.56$ are well described by the solution of the floating model given in Eq.~\eqref{eq:e4}.

%\textcolor{blue}{Please present the result on $t_\mathrm{max}$.
%This is important because one of the referees was interested in this result.
%Do not hide inconvenient results.
%}

%%%%%%%%%%%%%%%%%%%%%%%%%%%%%%%
\begin{figure}
\includegraphics[width=9cm]{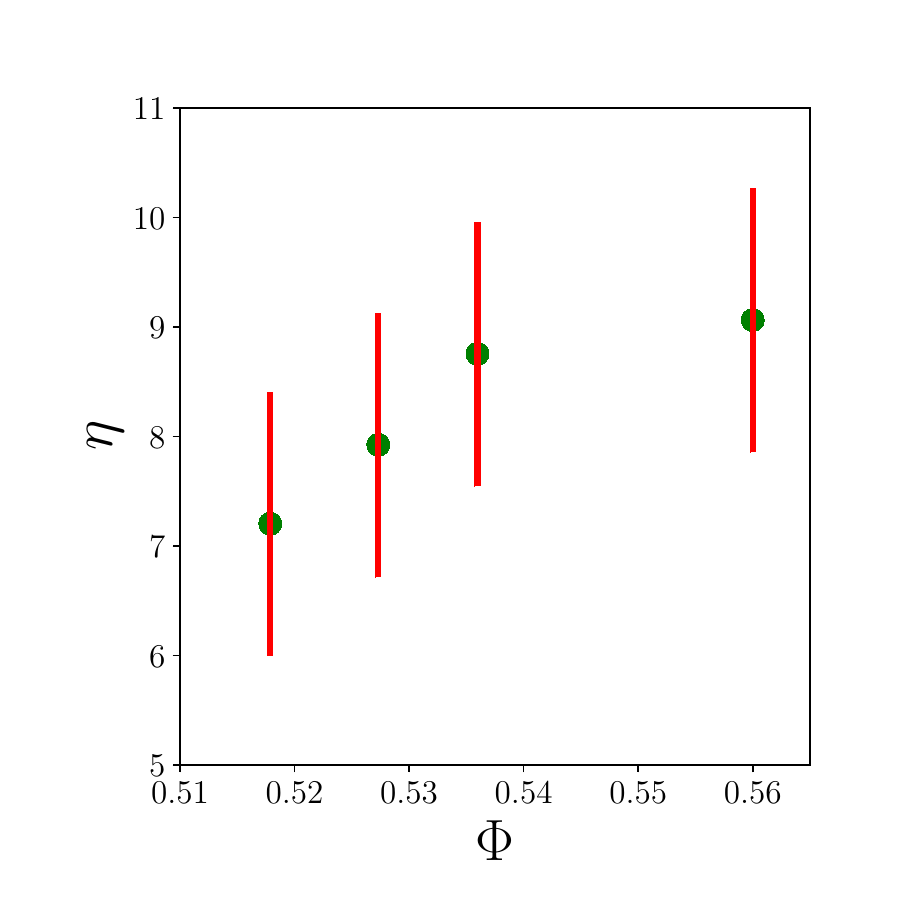}
\caption{Plot of the effective viscosity $\eta: = 3 \pi \eta_{{\rm eff}} a_{\rm I}^{3/2}/(g^{1/2}m_{\rm I})$ against $\Phi$. 
The red vertical lines express error bars of the estimated $\eta$.
%\textcolor{red}{LabelやLegendの字が変.pngを使っているせいだと思う。pngは使わないでepsかpdfにして下さい。また横軸のラベルは$\Phi$に変更して下さい。}
\label{fig:F5}}
\end{figure}

%%%%%%%%%%%%%%%%%%%%%%%%%

Figure \ref{fig:F5} exhibits how the dimensionless effective viscosity $\eta$ depends on $\Phi$.
The figure indicates that $\eta$ weakly depends on $\Phi$ in the range $\Phi=0.52 - 0.56$, where $\eta$ slightly increases with $\Phi$.
This result is unexpected because the viscosity in usual densely packed suspensions increases rapidly as the density approaches the jamming point or the random closed packing value. 
Since the fraction in the dynamical jammed region determines the effective viscosity, a lower $\Phi$ has a larger effective packing fraction.  
%This means that the range of the effective fraction is narrower than that presented in Fig.~\ref{fig:F5}.
We will discuss this insensitivity of $\eta$ to $\Phi$ in Sec.~\ref{sec:discussion}.
%This is consistent with Ref.~\cite{Pradipto23} where they observed only a slight increment of the effective viscosity after the impact of the projectile.

\section{Discussion}\label{sec:discussion}

As presented in the previous section, we found that the floating model provides a good description of our experimental results.
In particular, the behavior of the maximum drag force, $\tilde{F}_\mathrm{max}$, as a function of the impact speed $U_0$ shows reasonable agreement with the theoretical curves.
These results validate several simplifications and assumptions made in the floating model, namely, the approximation of buoyancy forces (see Appendix~\ref{appendix_b}), the neglect of elastic forces, and the use of Stokes drag when focusing on the early stage of the dynamics. %\textcolor{red}{for $t_\mathrm{half}<t<t_\mathrm{max}$.
It is important to note that the elastic force becomes significant during the late stage of the dynamics, although its contribution remains much smaller than that of the viscous drag force during the early stage of impact \cite{Brassard21, Pradipto21b, Pradipto23}.

Although several models have been proposed to describe impact processes in suspensions \cite{Waitukaitis12, Egawa19,Brassard21,Pradipto21a, Pradipto21b,Pradipto23}, most of these models include an elastic term to capture the rebound behavior observed in the late stage of the impact.
In other words, the elastic term strongly depends on the contact force between suspended particles, such as interparticle friction~\cite{Pradipto21a}.
In contrast, the floating model is particularly well-suited for describing the early-stage dynamics, where viscous effects dominate, and is insensitive to the contact force between particles.
Therefore, we conclude that the floating model accurately captures the early-stage dynamics in dense suspensions of millimeter-sized particles.

Surprisingly, the assumption of Stokes drag remains valid, even though it is typically justified only for suspensions of small particles with sufficiently low Reynolds numbers. Nevertheless, this assumption is self-consistent, as explained below.
We introduce an effective Reynolds number defined as $Re_\mathrm{eff}:=\rho_\mathrm{f} U_0  (d_\mathrm{max}+d_\mathrm{min})/(2\eta_\mathrm{eff})$.
Using our experimental parameters, $Re_\mathrm{eff}$ ranges from 0.47 to 4.44 at $\Phi = 0.54$, and from 0.58 to 5.37 at $\Phi = 0.56$, which are small enough to justify the use of the Stokes approximation.
These results support the use of the Stokes drag in the floating model. The apparent viscosity $\eta_\mathrm{a} := \eta_\mathrm{eff} / \eta_0$, where $\eta_0$ is the viscosity of water, ranges from 5675 to 6853.
These large values also support the Stokes approximation for the mean-field flow of the suspension liquid.
%\textcolor{red}{Can you write the apparent viscosity, which is the ratio of the effective viscosity to the solvent viscosity?}

%%%%%%%%%%%%%%%%%%%%%%%%%%%%%%%
\begin{figure}
\includegraphics[width=9cm]{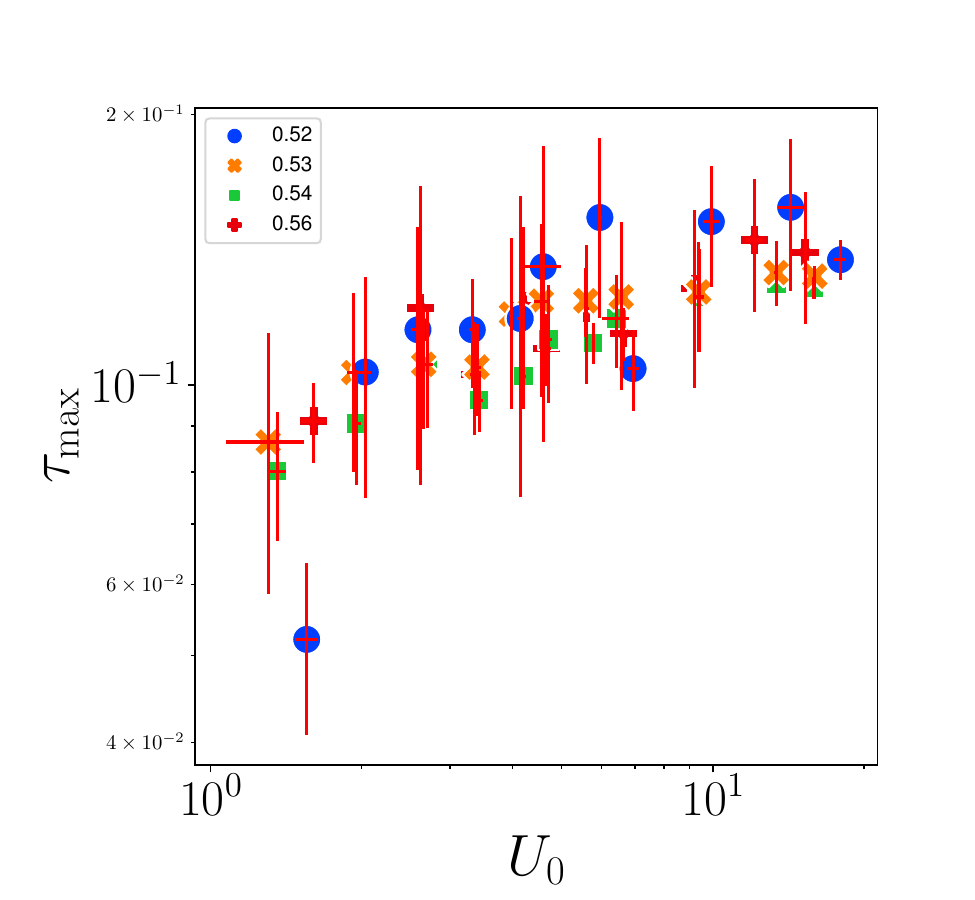}
\caption{The plots between $\tau_\mathrm{max}$ and $U_0$ for various $\Phi=0.52,0,53, 0.54$, and 0.56, where the vertical lines are error bars in the estimation of $\tau_\mathrm{max}$.
\label{fig:FC1}
}
\end{figure}

%%%%%%%%%%%%%%%%%%%%%%%%%

Unfortunately, our results for $\tau_\mathrm{max}$ is inconsistent with the prediction of the floating model~\cite{Pradipto21b}, where $\tau_\mathrm{max}$ weakly depends on $U_0$ (see Fig.~\ref{fig:FC1}), while the floating model predicts $\tau_\mathrm{max}\sim U_0^{-1/2}$ for large $U_0$. 
However, the previous experiment and simulation also do not get perfect agreement with the theoretical prediction~\cite{Brassard21,Pradipto21b}. 
This is partially because the floating model contains only one fitting parameter $\eta$ and has a quantitative deviation of the estimated $\tau_\mathrm{max}$ between the experiment and the theoretical one, as shown in Fig.~\ref{fig:F3}. 
More importantly, the onset of the drag force in the experiments may be governed by the mechanisms which is not involved in the floating model. 

%%%%%%%%%%%%%%%%%%%%%%%%%%%%%%%
\begin{figure}
\includegraphics[width=9cm]{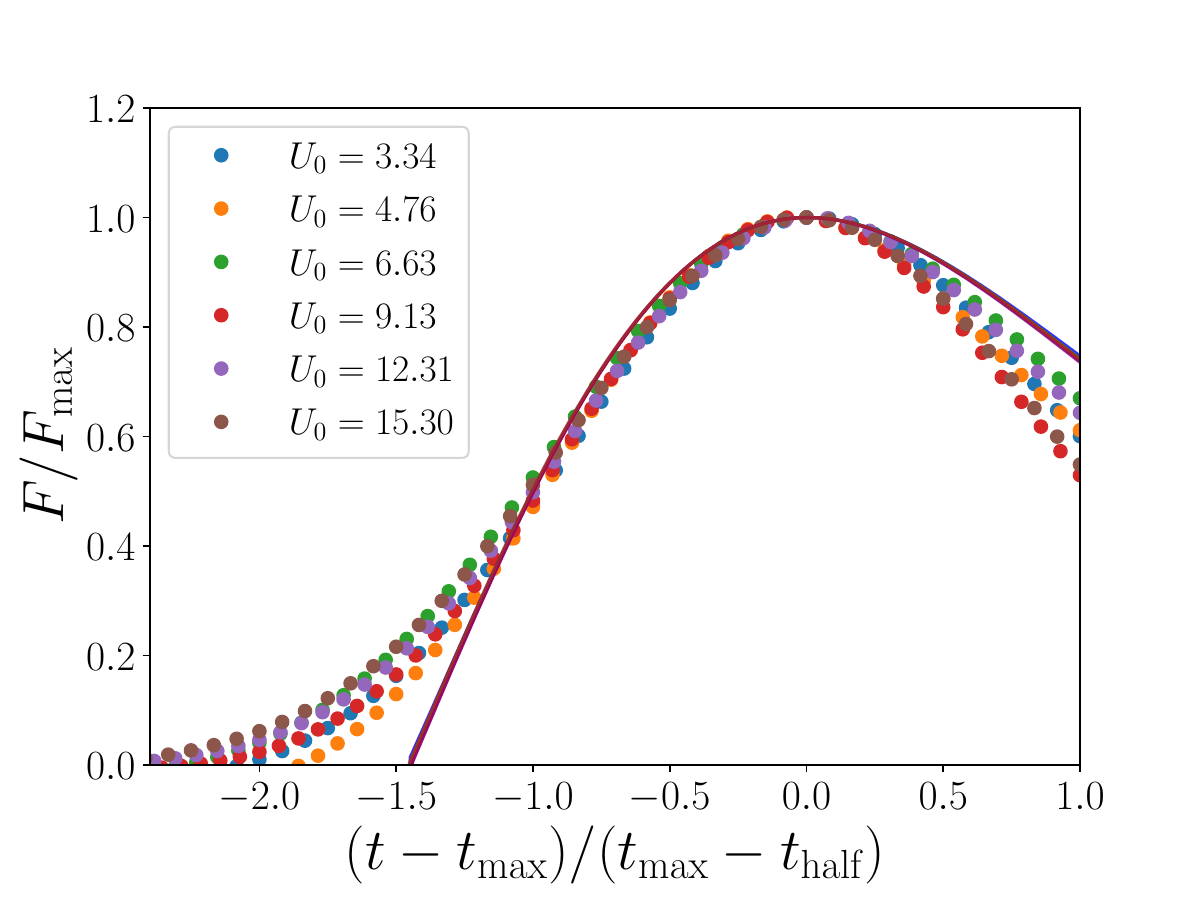}
\caption{
Scaling plots for normalized drag force $F/F_\mathrm{max}$ versus the dimensionless time $(t-t_\mathrm{max})/(t_\mathrm{max}-t_\mathrm{half})$ for various impact speeds at $\Phi=0.56$, where symbols show experimental observations and the solid lines are the solutions of the floating model for various $U_0$ with $\eta=9.06$ at the corresponding impact speeds.
\label{fig:F9}
}
\end{figure}

%%%%%%%%%%%%%%%%%%%%%%%%%

To clarify the time-dependent properties of the impact process, we introduce the half time, $t_\mathrm{half}$ to reach $F_\mathrm{max}/2$ where $\tau_\mathrm{half} = t_\mathrm{half} \sqrt{g/a_\mathrm{I}}$.\footnote{Similar to Fig.~\ref{fig:FC1}, the behavior of $\tau_\mathrm{half}$ versus $U_0$ cannot be described by the floating model.}
Figure~\ref{fig:F9} presents scaling plots of $F / F_\mathrm{max}$ versus $(t-t_\mathrm{max})/(t_\mathrm{max}-t_\mathrm{half})$ for various initial speeds $U_0$.
Experimental data for $(t-t_\mathrm{max})/(t_\mathrm{max}-t_\mathrm{half})<0$ and the solution of the floating model exhibit reasonable data collapse to support the scaling hypothesis near the peak.
Note that the elastic force plays an important role for $(t-t_\mathrm{max})/(t_\mathrm{max}-t_\mathrm{half})>0$.
%However, the curvature for the experiment slightly differs from that for the floating model. 
It should be noted that the scaling hypothesis for the experimental data is no longer valid near the point of impact of the projectile on the liquid, although the floating model satisfies the scaling for all time regions.
This indicates that another mechanism is necessary to explain Fig.~\ref{fig:FC1}. 

To evaluate sidewall effects due to the quasi-2D container used in the experiment, we also examine a cubic container (Appendix~\ref{appendix_a}).
The experimental results for $F_\mathrm{max}$ versus $U_0$ in the cubic container
(see Fig.\ref{fig:FA4}) are almost indistinguishable from those in Fig.\ref{fig:F4}.
This confirms the robustness of the floating model, regardless of the container shape.

We also examine several other situations in which the diameter of the projectile is set to be $D_\mathrm{I}=8$ mm instead of $D_\mathrm{I}=12$ mm as in the main text, and we have also examined a viscous liquid with 9 times larger viscosity using a water-glycerol suspension (see Appendix \ref{appendix_d}). 
Interestingly, we have confirmed that our results reported in the main text are unchanged even when we use smaller projectiles and/or viscous solvents.
These results suggest that our results in this paper are robust for relatively general situations.

As shown in Fig.~\ref{fig:F5}, the effective viscosity $\eta$ exhibits only weak dependence on the packing fraction $\Phi$.
This insensitivity can be attributed to two main factors.
First, the actual packing fractions are higher than the nominal ones due to imperfect density matching; for instance, $\varphi = 0.40$, 0.48, 0.51, and 0.56 correspond to $\Phi = 0.52$, 0.53, 0.54, and 0.56, respectively.
As a result, our experiments cover only a narrow range of $\Phi$.
Second, empirical relations suggest that the increase in effective viscosity is not substantial within this narrow range, especially under the assumption that the viscosity diverges at the random close packing fraction $\Phi_\mathrm{rcp} \approx 0.64$.
For example, using empirical relations for denser, sheared non-Brownian suspensions \cite{Boyer,Suzuki19}, the ratio is at most 2.0, although our system is not sheared.
If we apply the sedimentation theory\cite{HayakawaIchiki} used for $\Phi \le 0.50$, the ratio of effective viscosity between $\Phi = 0.56$ and $\Phi = 0.52$ is about 1.3, though this theory cannot be used for $\Phi>0.50$.
According to Appendix \ref{appendix_e}, the extended sedimentation theory of Ref.~\cite{HayakawaIchiki} for denser regions ($\Phi>0.50$) suggests that the ratio is merely 1.07.
Since the impact process is not sheared but is related to sedimentation, we should adopt the extended sedimentation theory to understand $\Phi$ dependence of $\eta$, which is insensitive to $\Phi$ in our observed region $0.52\le \Phi \le 0.56$,
These estimates suggest that the increase in effective viscosity over our observed range is not significant, assuming divergence at $\Phi_\mathrm{rcp}$, though a detailed quantitative justification remains an open question.

%%%%%%%%%%%%%%%%%%%%%%%%%%%%%%%%%%%%%%%%%%%%%%%%%%%%%%
\section{Conclusions}\label{sec:conclusions}
%{\it Concluding remarks}--.
%%%%%%%%%%%%%%%%%%%%%%%%%%%%%%%%%%%%%%%%%%%%%%%%%%%%%%

We have experimentally demonstrated that the floating model, originally developed for micrometer-sized particle suspensions, remains valid in describing the early-stage impact dynamics (for $t_\mathrm{half}<t<t_\mathrm{max}$) of millimeter-sized particle suspensions.
The observed scaling law $F_{\mathrm{max}} \sim u_0^{3/2}$ for $\eta U_0 \gg 1$ aligns with theoretical predictions, confirming the dominance of viscous drag in the initial phase of impact. 
Our findings extend the applicability of the viscous-based floating model and suggest that Stokes flow assumptions may hold under specific conditions, even for larger particles. 
This work bridges the gap between microscale simulations and macroscale experiments, offering new insights into the mechanics of dense suspensions. 

Looking ahead, a deeper understanding of why the floating model applies to suspensions of millimeter-sized particles remains an important direction for future research. 
Although this study primarily focuses on early-stage dynamics, future work should explore the role of elastic forces in the late stages of impact, so-called the IIH, as indicated by previous experimental~\cite{Egawa19} and theoretical~\cite{Pradipto21b,Pradipto23} studies.
The floating model fails to capture the elastic contribution at the instant of impact between the projectile and the particles in the suspension. Elucidating these dynamics remains an important direction for future work.
Also, using large-sized particles, we may visualize the force propagation in force chains in the suspended particles in the forthcoming study.

\vspace*{0.5cm}

\noindent
{\bf Acknowledgment}

We thank the useful comments by Fr\'{e}d\'{e}ric van Wijland, and Pradipto.
HM thanks Mahesh Bandi, and members of the Non-equilibrium Physics Unit at OIST, and Yutaro Motoori for their useful advice and comments.
This work is partially supported by  Grants-in-Aid of
MEXT, Japan for Scientific Research, Grant Nos. 
JP21H01006 and JP24K17022, and the Kyoto University Foundation. ChatGPT was used for writing assistance and program debugging.
%\textcolor{red}{Please add your KAKENHI.}
\vspace*{0.5cm}

\noindent
{\bf Data Availability}

The data that support the findings of this study are available from the corresponding author upon reasonable request.

\nocite{*}
%\bibliography{aipsamp}% Produces the bibliography via BibTeX.

%aipnum4-2.bst 2019-01-14 (MD) hand-edited version of apsrev4-1.bst
\begin{thebibliography}{0}%
\makeatletter
\providecommand \@ifxundefined [1]{%
 \@ifx{#1\undefined}
}%
\providecommand \@ifnum [1]{%
 \ifnum #1\expandafter \@firstoftwo
 \else \expandafter \@secondoftwo
 \fi
}%
\providecommand \@ifx [1]{%
 \ifx #1\expandafter \@firstoftwo
 \else \expandafter \@secondoftwo
 \fi
}%
\providecommand \natexlab [1]{#1}%
\providecommand \enquote  [1]{``#1''}%
\providecommand \bibnamefont  [1]{#1}%
\providecommand \bibfnamefont [1]{#1}%
\providecommand \citenamefont [1]{#1}%
\providecommand \href@noop [0]{\@secondoftwo}%
\providecommand \href [0]{\begingroup \@sanitize@url \@href}%
\providecommand \@href[1]{\@@startlink{#1}\@@href}%
\providecommand \@@href[1]{\endgroup#1\@@endlink}%
\providecommand \@sanitize@url [0]{\catcode `\\12\catcode `\$12\catcode
  `\&12\catcode `\#12\catcode `\^12\catcode `\_12\catcode `\%12\relax}%
\providecommand \@@startlink[1]{}%
\providecommand \@@endlink[0]{}%
\providecommand \url  [0]{\begingroup\@sanitize@url \@url }%
\providecommand \@url [1]{\endgroup\@href {#1}{\urlprefix }}%
\providecommand \urlprefix  [0]{URL }%
\providecommand \Eprint [0]{\href }%
\providecommand \doibase [0]{https://doi.org/}%
\providecommand \selectlanguage [0]{\@gobble}%
\providecommand \bibinfo  [0]{\@secondoftwo}%
\providecommand \bibfield  [0]{\@secondoftwo}%
\providecommand \translation [1]{[#1]}%
\providecommand \BibitemOpen [0]{}%
\providecommand \bibitemStop [0]{}%
\providecommand \bibitemNoStop [0]{.\EOS\space}%
\providecommand \EOS [0]{\spacefactor3000\relax}%
\providecommand \BibitemShut  [1]{\csname bibitem#1\endcsname}%
\let\auto@bib@innerbib\@empty
%</preamble>
\end{thebibliography}%


\begin{thebibliography}{999}

\bibitem{Russel89}
W. B. Russel, D. A. Saville, and W. R. Schowalter,  Colloidal Dispersions (Cambridge Univ. Press, Cambridge, 1989).

\bibitem{Hunter01}
R. J. Hunter, Foundations of colloid science (Oxford Univ. Press, Oxford, 2001).

\bibitem{Mewis11}
J. Mewis, N. J. Wagner, Colloidal Suspension Rheology (Cambridge University Press, New York, 2011).



\bibitem{Barnes89} H. A. Barnes, Shear‐Thickening (“Dilatancy”) in Suspensions of Nonaggregating Solid Particles Dispersed in Newtonian Liquids,
\href{https://doi.org/10.1122/1.550017}{J. Rheol. \textbf{33}, 329 (1989).}


\bibitem{Lootens05}
D. Lootens, H. van Damme, Y. H\'{e}mar, and P. H\'{e}braud, 
Dilatant Flow of Concentrated Suspensions of Rough Particles
\href{https://link.aps.org/doi/10.1103/PhysRevLett.95.268302}{Phys. Rev. Lett. \textbf{95}, 268302 (2005).}




\bibitem{Seto13}
R. Seto, R. Mari, J. F Morris, and M. M. Denn,
Discontinuous shear thickening of frictional hard-sphere suspensions,
\href{https://link.aps.org/doi/10.1103/PhysRevLett.111.218301}{Phys. Rev. Lett. \textbf{111}, 218301 (2013)}

\bibitem{Cwalina14}
C. D. Cwalina, and N. J. Wagner, Material properties of the shear-thickened state in concentrated near hard-sphere colloidal dispersions,
\href{https://doi.org/10.1122/1.4876935}{J. Rheol. \textbf{58}, 949 (2014).}


\bibitem{Brown14}
E. Brown and H. M. Jaeger, Shear thickening in concentrated suspensions:
Phenomenology, mechanisms and relations to jamming, 
\href{https://iopscience.iop.org/article/10.1088/0034-4885/77/4/046602}{Rep. Prog. Phys. \textbf{77},
046602 (2014)}.

\bibitem{Guy15}
B. M. Guy, M. Hermes, and W. C. K. Poon, Towards a Unified Description of the Rheology of Hard-Particle Suspensions, 
\href{https://link.aps.org/doi/10.1103/PhysRevLett.115.088304}{Phys. Rev. Lett. \textbf{115}, 088304 (2015).}


\bibitem{Ness22}
C. Ness, R. Seto, and R. Mari, The physics of dense suspensions,
\href{https://doi.org/10.1146/annurev-conmatphys-031620-105938}{Annu. Rev. Condens. Matter Phys. \textbf{13}, 97 (2022)}.

\bibitem{Brady88}
J. F. Brady and L. J. Durlofsky, The sedimentation rate of disordered suspensions,
\href{https://doi.org/10.1063/1.866808}{Phys. Fluids, \textbf{31}, 717 (1988).}

\bibitem{HayakawaIchiki}
H. Hayakawa and K. Ichiki, Statistical theory of sedimentation of disordered suspensions, \href{https://link.aps.org/doi/10.1103/PhysRevE.51.R3815}{Phys. Rev. E, {\bf 51}, R3815(R) (1995)}.

\bibitem{Boyer}
F. Boyer,  \'{E}. Guazzelli, and O. Pouliquen, Unifying Suspension and Granular Rheology, \href{https://link.aps.org/doi/10.1103/PhysRevLett.107.188301}{Phys. Rev. Lett. {\bf 107}, 188301 (2011)}.


\bibitem{Guazzelli}
\'{E}. Guazzelli, and O. Pouliquen, Rheology of dense granular suspensions,
\href{https://www.cambridge.org/core/journals/journal-of-fluid-mechanics/article/rheology-of-dense-granular-suspensions/4F792CE372121D52299422BAEADCDE74}
{J. Fluid Mech. \textbf{852} (2018)}.

\bibitem{Suzuki19}
K. Suzuki and H. Hayakawa, Theory for the rheology of dense non-Brownian suspensions: divergence of viscosities and $\mu-J$ rheology,
\href{https://doi.org/10.1017/jfm.2019.5}{J. Fluid Mech. \textbf{864}, 1125 (2019).}

\bibitem{Stronge} W. J. Stronge, Impact Mechanics (Cambridge Univ. Press, Cambridge, 2000).


\bibitem{Louge02} M. Y. Louge, and M. E. Adams, Anomalous behavior of normal kinematic restitution in the oblique impacts of a hard sphere on an elastoplastic plate, \href{https://link.aps.org/doi/10.1103/PhysRevE.65.021303}{Phys. Rev. E \textbf{65}, 021303 (2002).}

\bibitem{Kuninaka04} H. Kuninaka, and H. Hayakawa, Anomalous Behavior of the Coefficient of Normal Restitution in Oblique Impact, \href{https://link.aps.org/doi/10.1103/PhysRevLett.93.154301}{Phys. Rev. Lett. \textbf{93}, 154301 (2004).}

\bibitem{Falcon}
E. Falcon, C. Laproche, S. Fauve, C. Coste, Behavior of one inelastic ball bouncing repeatedly off the ground, \href{https://doi.org/10.1007/s100510050283}{Eur. Phys. J. B \textbf{3}, 45 (1998)}. 

\bibitem{Chastel}
T. Chastel, P. Gondret, A. Mongruel, Texture-driven elastohydrodynamic bouncing, \href{https://doi.org/10.1017/jfm.2016.580}{J. Fluid Mech. \textbf{805}, 577 (2016)}. 

\bibitem{Maruoka}
H. Maruoka, A framework for crossover of scaling law as a self-similar solution: dynamical impact of viscoelastic board, \href{https://doi.org/10.1140/epje/s10189-023-00292-9}{Eur. Phys. J. E \textbf{46}, 35 (2023)}.



\bibitem{Kuwabara87} G. Kuwabara and K. Kono, Restitution Coefficient in a Collision between Two Spheres, \href{https://iopscience.iop.org/article/10.1143/JJAP.26.1230}{Jpn. J. Appl. Phys. 26, 1230 (1987).}

\bibitem{Brilliantov96} N. V. Brilliantov, F. Spahn, J.-M. Hertzsch, and T. P\"{o}schel, Model for collisions in granular gases, \href{https://link.aps.org/doi/10.1103/PhysRevE.53.5382}{Phys. Rev. E \textbf{53}, 5382 (1996).}


\bibitem{Kuninaka09} H. Kuninaka, and H. Hayakawa, Simulation of cohesive head-on collisions of thermally activated nanoclusters, \href{https://link.aps.org/doi/10.1103/PhysRevE.79.031309}{Phys. Rev. E \textbf{79}, 031309 (2009)}.

\bibitem{Saitoh10} K. Saitoh, A. Bodrova, H. Hayakawa, and N. V. Brilliantov, Negative Normal Restitution Coefficient Found in Simulation of Nanocluster Collisions, \href{https://link.aps.org/doi/10.1103/PhysRevLett.105.238001}{Phys. Rev. Lett. \textbf{105}, 238001 (2010).}

\bibitem{Muller13} P. M\"{u}ller, M. Heckel, A. Sack, and T. P\"{o}schel, Complex Velocity Dependence of the Coefficient of Restitution of a Bouncing Ball, \href{https://link.aps.org/doi/10.1103/PhysRevLett.110.254301}{Phys. Rev. Lett. \textbf{110}, 254301 (2013).}

\bibitem{Uehara03} J. S. Uehara, M. A. Ambroso, R. P. Ojha, and D. J. Durian, Low-Speed Impact Craters in Loose Granular Media, \href{https://link.aps.org/doi/10.1103/PhysRevLett.90.194301}{Phys. Rev. Lett. \textbf{90}, 194301 (2003).}

\bibitem{Walsh03} Amanda M. Walsh, K. E. Holloway, P. Habdas, and J. R. de Bruyn, Morphology and Scaling of Impact Craters in Granular Media, \href{https://link.aps.org/doi/10.1103/PhysRevLett.91.104301}{Phys. Rev. Lett. \textbf{91}, 104301 (2003).}

\bibitem{Massimo04} M. Pica Ciamarra, A. H. Lara1, A. T. Lee1, D. I. Goldman, I. Vishik, and H. L. Swinney, Dynamics of Drag and Force Distributions for Projectile Impact in a Granular Medium, \href{https://link.aps.org/doi/10.1103/PhysRevLett.92.194301}{Phys. Rev. Lett. \textbf{92}, 194301 (2004).}

\bibitem{Lohse04} D. Lohse, R. Rauh\'{e}, R. Bergmann, and D. van der Meer, Creating a dry variety of quicksand, 
\href{https://doi.org/10.1038/432689a}{Nature \textbf{432}, 689 (2004). }

\bibitem{Katsuragi07} H. Katsuragi, and D. J. Durian, Unified force law for granular impact cratering, 
\href{https://doi.org/10.1038/nphys583}{Nature Phys. \textbf{3}, 420 (2007).}



\bibitem{Katsuragi16}
H. Katsuragi, \href{https://link.springer.com/book/10.1007/978-4-431-55648-0}{Physics of Soft Impact and Cratering} (Springer, Berlin, 2016).

\bibitem{Krizou20}
N. Krizou and A. H. Clark, 
Power-Law Scaling of Early-Stage Forces during Granular Impact,
\href{https://link.aps.org/doi/10.1103/PhysRevLett.124.178002}{Phys. Rev. Lett. \textbf{124}, 178002 (2020)}.



%\bibitem{Reis}
%N. Reis, C. Ainsley, B. Derby, Ink-jet delivery of particle suspensions by piezoelectric droplet ejectors, \href{https://doi.org/10.1063/1.1888026}{J. Appl. Phys. 97, 094903 (2005).}


%\bibitem{Passos}
%A. Passos, J. M. Sherwood, E. Kaliviotis, R. Agrawal, C. Pavesio, S. Balabani, The effect of deformability on the microscale flow behavior of red blood cell suspensions, \href{https://doi.org/10.1063/1.5111189}{Phys. Fluids {\bf 31}, 091903 (2019).}

%\bibitem{Aouae}
%O. Aouane, A. Scagliarini, J. Harting, Structure and rheology of suspensions of spherical strain-hardening capsules. \href{https://www.cambridge.org/core/journals/journal-of-fluid-mechanics/article/structure-and-rheology-of-suspensions-of-spherical-strainhardening-capsules/CBB9B2C7DD15E005B756B303F6A61C74}{J.  Fluid Mech., {\bf 911} (2021).}
    
%\end{comment}


\bibitem{Lee03}
Y. S. Lee, E. D. Wetzel, and N. J. Wagner, The ballistic impact characteristics
of Kevlar woven fabrics impregnated with a colloidal shear thickening fluid,
\href{https://doi.org/10.1023/A:1024424200221}{J. Mater. Sci. \textbf{38}, 2825 (2003)}.

\bibitem{Waitukaitis12}
S. R. Waitukaitis and H. M. Jaeger, Impact-activated solidification
of dense suspensions via dynamic jamming fronts, \href{https://doi.org/10.1038/nature11187}{Nature
(London) \textbf{487}, 205 (2012)}.

\bibitem{Waitukaitis13}
S. R. Waitukaitis,L.K.Roth, V. Vitelli, and H. M. Jaeger, Dynamic jamming fronts
\href{https://10.1209/0295-5075/102/44001}{EPL, \textbf{102}, 44001 (2013)} 


\bibitem{Han16}
E. Han, I. R. Peters, and H. M. Jaeger, High-speed ultrasound
imaging in dense suspensions reveals impact-activated solidification
due to dynamic shear jamming, 
\href{https://doi.org/10.1038/ncomms12243}{Nat. Commun. \textbf{7},
12243 (2016).}

\bibitem{Roche13}
M. Roche, E. Myftiu, M. C. Johnston, P. Kim, and H. A. Stone,
Dynamic Fracture of Nonglassy Suspensions, 
\href{https://link.aps.org/doi/10.1103/PhysRevLett.110.148304}{Phys. Rev. Lett.
\textbf{110}, 148304 (2013)}.

\bibitem{Maharjan18}
R. Maharjan, S. Mukhopadhyay, B. Allen, T. Storz, and E.
Brown, Constitutive relation for the system-spanning dynamically
jammed region in response to impact of cornstarch and
water suspensions, \href{https://link.aps.org/doi/10.1103/PhysRevE.97.052602}{Phys. Rev. E \textbf{97}, 052602 (2018)}.

\bibitem{Egawa19}
K. Egawa and H. Katsuragi, Bouncing of a projectile impacting
a dense potato-starch suspension layer, 
\href{https://doi.org/10.1063/1.5095678}{Phys. Fluids \textbf{31}, 053304
(2019)}.

\bibitem{Brassard21}
M. Brassard, N. Causley, N. Krizou, J. A. Dijksman, and A. H.
Clark, Viscous-like forces control the impact response of shearthickening dense suspensions, 
\href{https://doi.org/10.1017/jfm.2021.611}{J. Fluid Mech. \textbf{923}, A38 (2021)}.


\bibitem{Pradipto21a}
Pradipto and H. Hayakawa, Impact-induced hardening in dense
frictional suspensions, 
\href{https://link.aps.org/doi/10.1103/PhysRevFluids.6.033301}{Phys. Rev. Fluids \textbf{6}, 033301 (2021)}.
\bibitem{Pradipto21b}
Pradipto and H. Hayakawa, Viscoelastic response of impact
process on dense suspensions, \href{https://doi.org/10.1063/5.0061196}{Phys. Fluids \textbf{33}, 093110 (2021)}.

\bibitem{Pradipto23}
Pradipto and H. Hayakawa, Effective viscosity and elasticity in dense suspensions under impact: Toward a modeling of walking on suspensions,
\href{https://link.aps.org/doi/10.1103/PhysRevE.108.024604}{Phys. Rev. E \textbf{108}, 024604 (2023)}.


\bibitem{Kann11} S. von Kann, J. H. Snoeijer, D. Lohse, and D. van der Meer, Nonmonotonic settling of a sphere in a cornstarch suspension, \href{https://link.aps.org/doi/10.1103/PhysRevE.84.060401}{Phys.Rev.E \textbf{84}, 060401(R) (2011).}

\bibitem{Maharjan17} R. Maharjan and E. Brown, Giant deviation of a relaxation time from generalized Newtonian theory in discontinuous shear thickening suspensions, \href{https://link.aps.org/doi/10.1103/PhysRevFluids.2.123301}{Phys. Rev. Fluids \textbf{2}, 123301 (2017).}

\bibitem{Cho22}  J. H. Cho, A. H. Griese, I. R. Peters, and I. Bischofberger, Lasting effects of discontinuous shear thickening in cornstarch suspensions upon flow cessation, \href{https://link.aps.org/doi/10.1103/PhysRevFluids.7.063302}{Phys. Rev. Fluids \textbf{7}, 063302 (2022).}

\bibitem{Barik22} S. Barik and S. Majumdar, Origin of Two Distinct Stress Relaxation Regimes in Shear Jammed Dense Suspensions, \href{https://link.aps.org/doi/10.1103/PhysRevLett.128.258002}{Phys. Rev. Lett. \textbf{128}, 258002 (2022).}


\bibitem{Otsuki11}
M. Otsuki and H. Hayakawa, 
Critical scaling near jamming transition for frictional granular particles,
\href{https://link.aps.org/doi/10.1103/PhysRevE.83.051301}{Phys. Rev. E \textbf{83}, 051301 (2011).}



\bibitem{Johnson}
K. L. Johnson, \href{https://doi.org/10.1017/CBO9781139171731}{ Contact Mechanics (Cambridge University Press, Cambridge, 1985)}. 

\bibitem{Mandal24}
M. K. Mandal, and S. Roy, 
High speed impact on granular media: breakdown of conventional inertial drag models
\href{https://pubs.rsc.org/en/content/articlelanding/2024/sm/d3sm01410j}{Soft Matter, \textbf{20}, 877 (2024).}

\bibitem{Mitarai}
N. Mitarai, and F. Nori, Wet granular materials, \href{https://doi.org/10.1080/00018730600626065}{Advances in Phys., {\bf 55}, 1--45 (2006). }




\bibitem{PY}
J. K. Percus and G. J. Yevick, { Analysis of Classical Statistical Mechanics by Means of Collective Coordinates},
\href{https://link.aps.org/doi/10.1103/PhysRev.110.1}{Phys. Rev. \textbf{110}, 1 (1958)}.


 \bibitem{Torquato}   
    S. Torquato,
    { Nearest-neighbor statistics for packings of hard spheres and disks},
    \href{https://doi.org/10.1103/PhysRevE.51.3170}
    {Phys. Rev. E \textbf{51}, 3170 (1995)}.

\bibitem{CS}
    N. F. Carnahan and K. E. Starling,
    { Equation of State for Nonattracting Rigid Spheres},
    \href{https://doi.org/10.1063/1.1672048}
    {J. Chem. Phys. \textbf{51}, 635 (1969)}.


\end{thebibliography}

\appendix

\section{The dynamical impact of experiments on the cubic cell}\label{appendix_a}

\begin{figure}
\includegraphics[width=9cm]{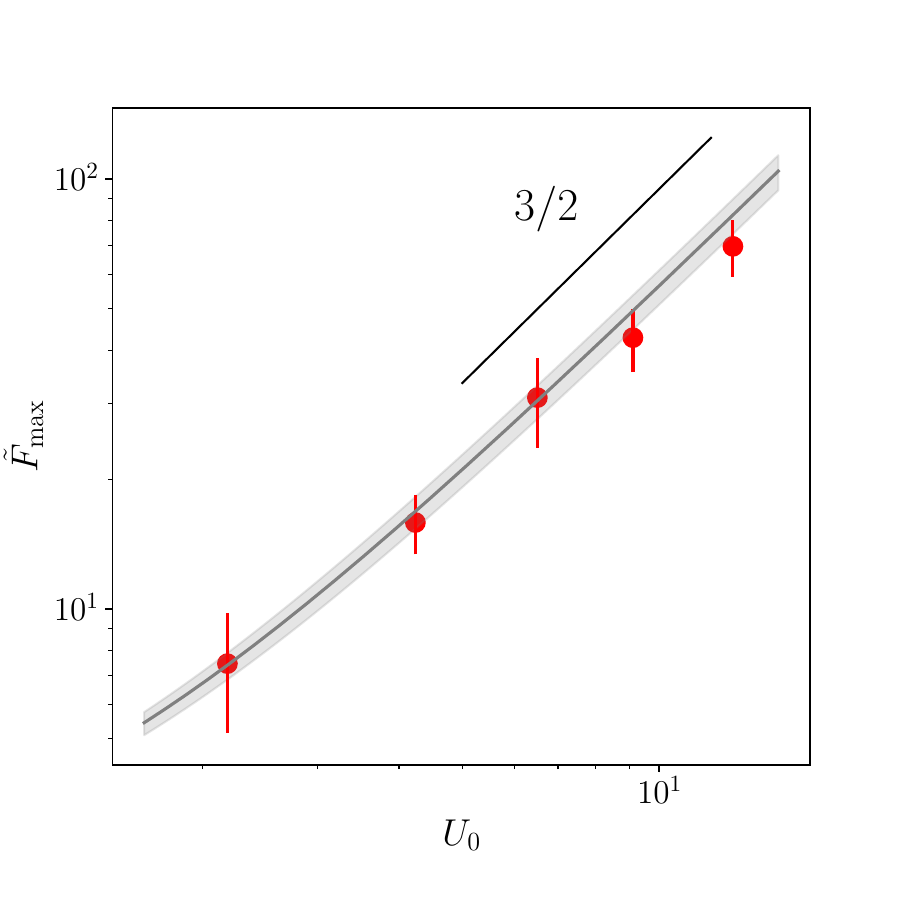}
\caption{The plots between the dimensionless maximum force $\tilde{F}_\mathrm{max}:=F_\mathrm{max}/m_{\rm I}g$ and impact-speed $U_0:=u_0/ \sqrt{a_{\rm I} g}$ for the cubic cell in the packing fraction $\Phi = 0.56$. 
The red solid circles indicate the experimental results and the black solid lines are obtained by the numerical solutions of the floating model Eq.~\eqref{eq:e4}, where $\eta=10.09 \pm 1.90$. The gray area around the solid line indicates the possible solutions within the errors of the estimated $\eta$.  
The black solid line segments in large $U_0$ are the guide lines $\tilde{F}_\mathrm{max}\propto U_0^{3/2}$ as expected from Eq.~\eqref{eq:e5a}.
%\textcolor{red}{LabelやLegendの字が変.pngを使っているせいだと思う。pngは使わないでepsかpdfにして下さい。}
\label{fig:FA4}}
\end{figure}

In this appendix, we examine a cube container ($W$, $D$, and $H$ are, respectively, given by 94 mm, 94 mm, and 90 mm) to evaluate the possible influence of the side wall effects from the quasi-2D container used in the main text (see Fig.~\ref{fig:FA4}). 
We use the packing fraction $\Phi = 0.56$ using $N = 1197$.
The resulting relationship of $F_\mathrm{max}$ versus $U_0$ behavior closely matches that observed in Fig.~\ref{fig:F4}.
Thus, we conclude the robustness of the floating model regardless of container shape or boundary effects.

%%%%%%%%%%%%%%%%%%%%%%%%%%%%%%%

%%%%%%%%%%%%%%%%%%%%
\section{The expression of the buoyancy force}\label{appendix_b}
%%%%%%%%%%%%%%%%%

If a projectile with the density $\rho_{\rm I}$ is located in a fluid with density $\rho_{\rm f}(\ne \rho_{\rm I})$, the motion of the projectile is affected by the buoyancy force. 
The expression of the buoyancy force must depend on the immersed volume of the projectile in the fluid,
where the immersed volume $V(z)$ of the projectile at the location $z$ (see Fig.~2 in the main text) is expressed as
\begin{eqnarray}
V\left( z \right)  &=& \int _{0}^{z} \pi {a_{\rm I}^{\prime}}^2 d z^{\prime} \nonumber \\ 
    &=& \frac{\pi a_{\rm I} ^3}{3}\left[3\left(\frac{z}{a_{\rm I}} \right)^2 - \left(\frac{ | z |}{a_{\rm I}} \right)^3\right]. \label{eq:e6a}
\end{eqnarray}
This expression is valid for $-2 a_{\rm I} \leq z \leq 0$.

%%%%%%%%%%%%%%%%%%%%%%%%%%%%%%%
\begin{figure}
\includegraphics[width=9cm]{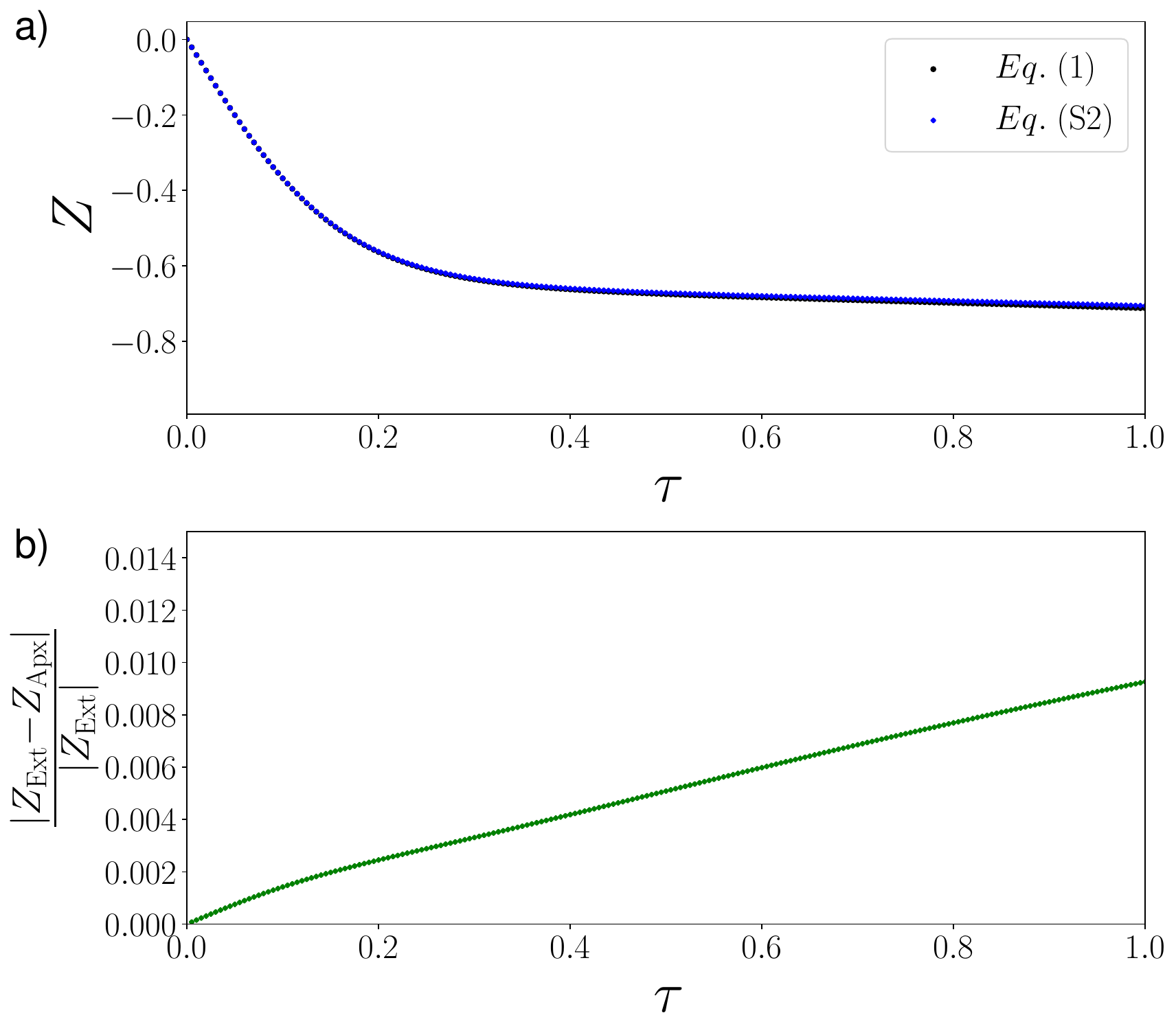}
\caption{Comparison between the numerical solutions of the exact expression of the buoyancy force Eq.~\eqref{eq:e6} and the approximate expression Eq.~(1), where we plot the time evolution of position a) and their difference of two expressions b). \label{fig:FA1}
%\textcolor{red}{Fonts of labels are too small.
%\textcolor{red}{too small fonts.}
}
\end{figure}

%%%%%%%%%%%%%%%%%%%%%%%%%

Thus, the buoyancy force is given by
\begin{eqnarray}\label{eq:e6}
F_{\rm B} &=&-g\{ \rho_{\rm I} V_\mathrm{I}  - \rho_{\rm f} V\left(z \right)\}     \nonumber \\
 &=& -m_{\rm I} g \left\{1-\frac{\rho_{\rm f}}{4\ \rho_{\rm I}}\left[3\left( \frac{z}{a_{\rm I}}\right)^2 -\left( \frac{|z|}{a_{\rm I}}\right)^3 \right] \right\} ,
\end{eqnarray}
where $\rho_{\rm f}$ is the density of fluid and $\rho_{\rm I}$ is the density of the projectile.
Once we adopt Eq.~\eqref{eq:e6}, we cannot solve the floating model exactly.
Fortunately, if we assume $\rho_{\rm I}\gg \rho_{\rm f}$, the simplified buoyancy force Eq.~(1) as in Refs.~\cite{Pradipto21b,Pradipto23} 
%\begin{equation}
%F_{\rm B} \simeq  -m_{\rm I} g \left(1 - \frac{\rho_{\rm f}}{\rho_{\rm I}} \right)
%\label{eq:e7}
%\end{equation}
gives the reasonable agreement with that in Eq.~\eqref{eq:e6} as shown in Fig.~\ref{fig:FA1} a).
The error caused by the approximation $Z_\mathrm{Apx}$ in Eq. (1) plots $|Z_\mathrm{Ext}-Z_\mathrm{Apx}|/|Z_\mathrm{Ext}|$ in Fig.~\ref{fig:FA1} b), where $Z_\mathrm{Ext}$ is the solution of the floating model (4) with Eq.~\eqref{eq:e6}.
As shown in this figure, the error by the simplified expression in Eq. (1) is less than 0.8$\%$. 
Thus, for simplicity, we adopt Eq.~(1) for the buoyancy force in this paper.

%%%%%%%%%%%%%%%%%%%%%%%
\section{Exact solution of floating model}\label{appendix_c}
%%%%%%%%%%%%%%%%%%%%%%

As shown in Refs.~\cite{Pradipto21b,Pradipto23}, the floating model Eq. (4) with Eq. (1) can be solved exactly.
The explicit expression of $Z(\tau)$ is given by
\begin{equation}
Z\left(\tau \right) = -\left[\frac{4\left(1-\xi \right)}{\eta^2} \right]^{1/3} \frac{-{\rm Ai}^{\prime}\left(\Xi \right){\rm Bi}^{\prime}\left(\Lambda \right)+{\rm Ai}^{\prime}\left(\Lambda \right){\rm Bi}^{\prime}\left(\Xi \right)}{{\rm Ai}^{\prime}\left(\Lambda \right){\rm Bi}\left(\Xi \right)+{\rm Ai}\left(\Xi \right){\rm Bi}^{\prime}\left(\Lambda \right)}
,
\label{eq:e8}
\end{equation}
where $\Xi: = \left(\frac{\eta\left(1-\xi\right)}{2} \right)^{1/3}\left(\tau + \frac{U_0}{1-\xi}\right)$, $
\Lambda: = \left(\frac{\eta}{2\left(1-\xi\right)^2} \right)^{1/3}U_0$. ${\rm Ai}\left(x \right)$ and ${\rm Bi}\left(x \right)$ are Airy functions as
${\rm Ai}\left(x \right):= \frac{1}{\pi} \int_{0}^{\infty}\cos \left(\frac{t^3}{3}+xt \right) dt $, and 
${\rm Bi}\left(x \right):= \frac{1}{\pi} \int_{0}^{\infty}\left[ e^{-\frac{t^3}{3}+xt} +\sin \left( \frac{t^3}{3} +xt\right)\right] dt$, ${\rm Ai}^{\prime}(x)$ and ${\rm Bi}^{\prime}(x)$ are their derivatives with respect to $x$, respectively.
Note that $F_\mathrm{max}$ cannot be obtained exactly, although the asymptotic expression for $U_0\gg 1$ can be evaluated~\cite{Pradipto21b}.

%%%%%%%%%%%%%%%%%%%%%%%
%\section{The plots between $\tilde{F}_\mathrm{max}$ and $U_0/\tau_\mathrm{half}$ }\label{appendix_d1}
%%%%%%%%%%%%%%%%%%%%%%

%%%%%%%%%%%%%%%%%%%%%%%
\section{Comparison between various solution viscosities and projectile sizes.}\label{appendix_d}
%%%%%%%%%%%%%%%%%%%%%%

\begin{figure}
\includegraphics[width=9cm]{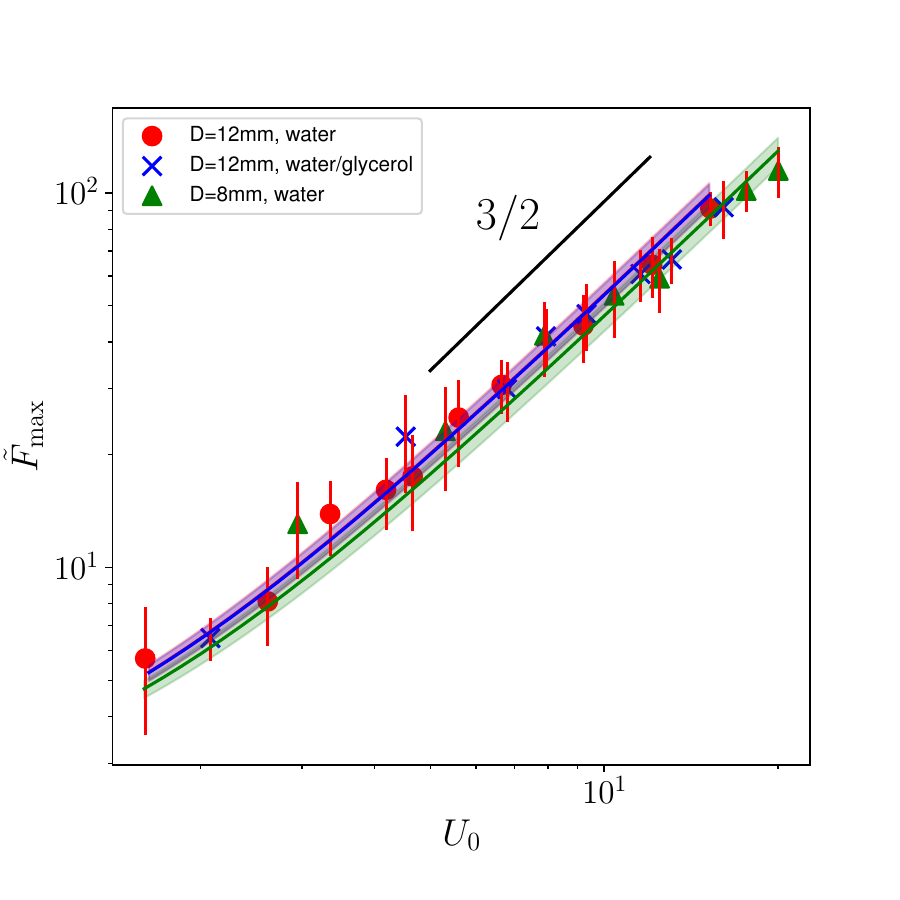}
\caption{A comparison of the $\tilde{F}_\mathrm{max}$--$U_0$ plots for the water suspension with $D = 12~\mathrm{mm}$, the water/glycerol suspension with $D = 12~\mathrm{mm}$, and water suspension with $D = 8~\mathrm{mm}$. All the packing fraction is $\Phi=0.56$.  \label{fig:F7}
}
\end{figure}

We compare the plots of $F_\mathrm{max}$ versus $U_0$ using a water-glycerol suspension, and a smaller projectile in Fig.~\ref{fig:F7}. 
The suspension solution is prepared by mixing with water and glycerol (water:glycerol=1:1 in volume fraction), whose viscosity is increased to be about 9 times larger than water, $\eta_0 =8.8~{\rm mPa~s}$. 
Then the glycerol/water solution is mixed with particles with the packing fraction $\Phi=0.56$. 
As the density of the glycerol/water solution becomes $1140~{\rm kg/m^3}$, the density matching is impossible to achieve in this case. 
In a separate experiment, a smaller projectile with a diameter $D = 8~\mathrm{mm}$ is used in a water-based suspension with the same packing fraction ($\Phi = 0.56$). 
Remarkably, all the data points collapse onto a single curve, exhibiting a power-law scaling with an exponent of $3/2$ at higher impact speeds.

The effective viscosity is estimated as $\eta = 8.99 \pm 1.30$ for the suspension of water/glycerol, $\eta = 6.88 \pm 1.23$ for a smaller projectile $D=8~{\rm mm}$. 
In comparison with water suspension, $\eta=9.06 \pm 1.45$, the data plots are seemingly overlapping. 
Thus, the floating model can be used for suspensions with smaller projectiles and/or viscous solvents.

%%%%%%%%%%%%%%%%%%%%%%%%%%%%%%%

%%%%%%%%%%%%%%%%%%%%%%%%%

%%%%%%%
\section{Extended sedimentation theory}\label{appendix_e}
%%%%%%%

In this appendix, we attempt to extend the sedimentation theory for $\Phi < 0.49$~\cite{HayakawaIchiki} to the denser situation, $\Phi > 0.49$, using an empirical argument.
The sedimentation rate for one particle sedimentation rate $U_0^\mathrm{sed}$, in a fluid is expressed as 
\begin{equation}
U^\mathrm{sed}_0:=\frac{2}{9\eta_0}\Delta \rho g a_\mathrm{I}^2  ,  
\end{equation}
where $\eta_0$ is the viscosity of the solvent, $\Delta \rho:=\rho_\mathrm{I}-\rho_\mathrm{f}$ and $a_\mathrm{I}:=d_\mathrm{I}/2$.
If we are interested in the sedimentation rate in random suspensions with the packing fraction $\Phi$, this can be written as
\begin{equation}
    \frac{U^\mathrm{sed}(\Phi)}{U_0^\mathrm{sed}}=\frac{\eta_0}{\eta_\mathrm{eff}} \propto \eta^{-1}.
\end{equation}
Therefore, to estimate the viscosity in the dense random suspensions is to estimate $U_0^\mathrm{sed}(\Phi)$.
It is known that the sedimentation rate is the evaluation of the ensemble average of the mobility matrix $M$, i.e. 
\begin{equation}
    \langle M \rangle= \frac{U^\mathrm{sed}(\Phi)}{U_0^\mathrm{sed}}\propto \eta^{-1}.
\end{equation}

The sedimentation theory in Ref.~\cite{HayakawaIchiki} consists of two parts: one is from the long-range Rotne-Prager part, and the other is the lubrication part.
Both contributions contain parts proportional to the density and the convolution integrals containing the pair correlation function. 
The lubrication part is unaffected by the density.
Thus, the averaged lubrication matrix over random suspensions $\langle \mathsf{R}^\mathrm{lub}\rangle$
can be evaluated as $\langle \mathsf{R}^\mathrm{lub} \rangle\approx 1.492 \Phi$.~\cite{HayakawaIchiki}

On the other hand, Brady and Durlofsky~\cite{Brady88} evaluated the long-range part, the random average of the mobility matrix $\langle M^\infty \rangle$ as
\begin{equation}\label{Brady}
\langle M^\infty \rangle=1-5\Phi -\frac{\Phi^2}{5}+3\Phi \int_{2}^\infty dr' r'\{g_2(r')-1\} ,
\end{equation}
where $g_2(r)$ is the pair-correlation function with the dimensionless distance $r$ normalized by the radius of each particle.
If we adopt the Percus-Yevick approximation~\cite{PY}, the integral in Eq.~\eqref{Brady} can be evaluated as 
$\int_2^\infty dr r(g_2^\mathrm{PY}(r)-1)= -2(5-\Phi+\Phi^2/2)/5(1+2\Phi)$. 
Thus, we obtain $\langle M^\infty \rangle\approx (1-\Phi)^3/(1+2\Phi)$, which gives a reasonable sedimentation rate with the approximation~\cite{Brady88}:
$\langle M\rangle \approx (\langle M^\infty \rangle^{-1}+\langle R^\mathrm{lub}\rangle)^{-1} \approx (1-\Phi)^3/(1+2\Phi+1.492 \Phi(1-\Phi)^3)$ for $\Phi<0.49$.
However, the calculation of $g_2(r)$ for $\Phi>0.49$ is difficult.
Instead, we adopt the following empirical relation.
It is known that $g_2(r)$ is zero for $r<2$, has a sharp peak at $r=2$, and oscillates around $1$ for $r>3$ for dense liquids.
Thus, we assume the integral part is proportional to the radial distribution at contact as
\begin{equation}\label{empirical}
\int_2^\infty dr r(g_2(r)-1) \approx \frac{g_0^\mathrm{T}(\Phi)}{g_0^\mathrm{CS}(\Phi)}\left\{ \frac{\Phi(11-\Phi)}{5(1+2\Phi)}\right\},
\end{equation}
where $g_0^\mathrm{T}(\Phi)=(1-\Phi_A/2)(\Phi_\mathrm{rcp}-\Phi_A)/\{(1-\Phi_A)^3(\Phi_\mathrm{rcp}-\Phi\}$ introduced in Ref.~\cite{Torquato} and $g_0^\mathrm{CS}(\Phi)=(1-\Phi/2)/(1-\Phi)^3$ introduced in Ref.~\cite{CS} with $\Phi_A=0.49$ and $\Phi_\mathrm{rcp}=0.639$.
Substituting Eq.~\eqref{empirical} into Eq.~\eqref{Brady}, we obtain the empirical far-field contribution to the sedimentation rate for $\Phi>0.49$ as
\begin{equation}\label{E6}
\langle M^\infty \rangle\approx 1+\Phi-\frac{\Phi^2}{5}+\frac{g_0^\mathrm{T}(\Phi)}{g_0^\mathrm{CS}(\Phi)}\left\{ \frac{\Phi(11-\Phi)}{5(1+2\Phi)}\right\}. 
\end{equation}
Then, the sedimentation rate can be obtained as 
\begin{align}\label{E7}
\langle M\rangle \approx (\langle M^\infty \rangle^{-1}+\langle R^\mathrm{lub}\rangle)^{-1}.
\end{align}
If we believe Eqs.~\eqref{E6} and \eqref{E7}, $\langle M \rangle \approx 2.8$ for $\Phi=0.52$ and $\langle M \rangle \approx 3.0$ for $\Phi=0.56$.
This is an interesting result that supports the insensitivity of $\eta$ against $\Phi$.

%%%%%%%%%%%%%%%%%%%%%%%%%%

\end{document}